\documentclass[a4paper,fleqn,usenatbib]{mnras}

\usepackage{natbib}

\usepackage{newtxtext,newtxmath}
\usepackage[T1]{fontenc}
\usepackage{ae,aecompl}

\usepackage{todonotes}
\usepackage{color}

\usepackage{graphicx}	% Including figure files
\usepackage{amsmath}	% Advanced maths commands
\usepackage{amssymb}	% Extra maths symbols

\usepackage{natbib}
\newcommand{\lcdm}{\Lambda\mathrm{CDM}}
\newcommand{\avg}[1]{\left\langle #1 \right\rangle}
\newcommand{\rmem}{\left\langle R_{\mathrm{mem}} \right\rangle}
\newcommand{\rlam}{R_{\lambda}}
\newcommand{\pmem}{p_{\mathrm{mem}}}
\newcommand{\rmiya}{\langle R^{\mathrm{M16}}_{\mathrm{mem}} \rangle}
\newcommand{\smiya}{\tilde{R}^{\mathrm{M16}}_{\mathrm{mem}}}
\newcommand{\rzu}{\langle R^{\mathrm{cut}}_{\mathrm{mem}} \rangle}
\newcommand{\smem}{\tilde{R}_{\mathrm{mem}}}
\newcommand{\dd}{\mathrm{d}}
\newcommand{\hmsol}{h^{-1}M_\odot}

\newcommand{\hmpc}{h^{-1}\mathrm{Mpc}}
\newcommand{\hgpc}{h^{-1}\mathrm{Gpc}}
\newcommand{\cubichgpc}{h^{-3}\mathrm{Gpc}^3}

\newcommand{\hhmsol}{h^{-2}M_\odot}

\newcommand\redmapper{redMaPPer}
\newcommand\bigmdpl{\texttt{BigMDPL}}
\newcommand\rockstar{\texttt{ROCKSTAR}}
\newcommand\ihod{\texttt{iHOD}}

\title[halo assembly bias vs.\ projection effect]{On the Level of Cluster Assembly Bias in SDSS}

\author[Zu et al. 2016]{Ying  Zu$^{1, 2}$\thanks{E-mail: zu.4@osu.edu},
Rachel Mandelbaum$^{2}$,
Melanie Simet$^{2}$,
Eduardo Rozo$^{3}$,
Eli S. Rykoff$\hspace{0.1em}^{4, 5}$
\\
$^{1}$Center for Cosmology and AstroParticle Physics (CCAPP),
Ohio State University, Columbus, OH 43210, USA\\
$^{2}$McWilliams Center for Cosmology, Department of Physics, Carnegie Mellon University, 5000 Forbes Avenue,
Pittsburgh, PA 15213, USA\\
$^{3}$Department of Physics, University of Arizona, Tucson, AZ 85721, USA\\
$^{4}$Kavli Institute for Particle Astrophysics and Cosmology, Department of Physics, Stanford University,
Stanford, CA 94305, USA\\
$^{5}$SLAC National Accelerator Laboratory, Menlo Park, CA 94025, USA\\
}

\date{Accepted XXX. Received YYY; in original form ZZZ}

\pubyear{2016}

\begin{document}
\label{firstpage}
\pagerange{\pageref{firstpage}--\pageref{lastpage}}
\maketitle

\begin{abstract}
  Recently, several studies have discovered a strong discrepancy between the large-scale clustering biases of
  two subsamples of galaxy clusters at the same halo mass, split by their average projected membership
  distances $\rmem$. The level of this discrepancy significantly exceeds the maximum halo assembly bias
  predicted by $\lcdm$. We explore whether some of the large-scale bias differences could be caused by
  projection effects in $\rmem$ due to other systems along the line-of-sight.  We thoroughly investigate the
  assembly bias of the \redmapper\ clusters in SDSS, by defining a new variant of the average membership
  distance estimator $\smem$ that is robust against projection effects in the cluster membership
  identification. Using the angular mark correlation functions, we show that the large-scale bias differences
  when splitting by $\rmem$ can be mostly attributed to projection effects. After splitting by $\smem$, the
  anomalously large signal is reduced, giving a ratio of $1.02\pm0.14$ between the two clustering biases as
  measured from weak lensing.  Using a realistic mock cluster catalogue, we predict that the bias ratio
  between two $\smem$-split
  subsamples should be ${\simeq}1.10$, which is ${>}60\%$ weaker than the maximum halo assembly
  bias~($1.24$) when split by halo concentration. Therefore, our results demonstrate that the level of halo
  assembly bias exhibited by clusters in SDSS is consistent with the $\lcdm$ prediction. With a ten-fold
  increase in cluster numbers, deeper ongoing surveys will enable a more robust detection of halo assembly
  bias. Our findings also have important implications for quantifying the impact of projection effects on
  cosmological constraints using photometrically-selected clusters.
\end{abstract}

%%%%%%%%%%%%%%%%%%%%%%%%%%%%%%%%%%%%%%%%%%%%%%%%%%

%%%%%%%%%%%%%%%%% BODY OF PAPER %%%%%%%%%%%%%%%%%%

%%%%%%%%%%%%%%%%%%%%%%%%%%%%%%%%%%%%%%%%%%%%%%%%%%
\section{Introduction}
\label{sec:intro}

The cold dark matter~(CDM) structure formation theory predicts that the large-scale bias of halo clustering
relative to the dark matter depends not only on halo mass, but also on other intrinsic halo properties such as
concentration, formation time, substructure abundance, and spin~\citep{sheth2004, gao2005, wechsler06,
harker2006, jing2007, li2008}. In particular, above the characteristic non-linear mass scale,
high-concentration halos have a lower clustering bias than their low-concentration counterparts, but for lower
mass halos the trend is reversed --- more concentrated halos exhibit higher clustering biases. This extra
dependence of halo bias on properties other than halo mass, often referred to as ``halo assembly
bias''~\citep{gao2007}, could be an important source of theoretical systematic uncertainty in the cosmological
constraints~\citep{wu2008, mcewen2016}, and potentially leave imprints on the formation and distribution of
galaxies~\citep{berlind2006, zhu2006, weinmann2006, blanton2007, croton2007, zu2008, deason2013,
kauffmann2013, zentner2014, lehmann2015, paranjape2015, zentner2016}. Therefore, it is a vital task to
directly detect halo assembly bias in observations and explore whether the observed assembly bias signal is
consistent with theoretical expectations from $\lcdm$.

The assembly bias phenomena in low and high-mass regimes have very distinct
theoretical origins. For low-mass halos, the assembly bias effect is mainly
caused by the tidal heating and stripping of old halos in dense environments or
by nearby larger systems, which suppressed the growth they would have otherwise
experienced in the field~\citep{diemand2007, hahn2009}. Additionally, when a
bound group of subhalos was tidally disrupted after entering into a massive
halo, a fraction of the subhalos would be ejected in highly-eccentric orbits and
contribute to the overall assembly bias after they became distinct
halos~\citep{ludlow2009, wang2009}. For the very massive halos that we will
focus on in this paper, \citet{dalal08} demonstrated that their assembly bias is
directly related to the curvatures of Lagrangian peaks in the initial Gaussian
random density field, analogous to the connection between linear halo bias and
peak height in the peak background-split formalism~\citep{bardeen1986,
sheth1999}. In a nutshell, two rare peaks of the same height but different
curvatures usually appeared in different large-scale environments, and
subsequently collapsed into two equal-mass clusters with different
concentrations.

Detecting halo assembly bias signal in observations requires a robust halo
finder, accurate measurements of halo mass, and a good proxy for halo
concentration or formation time~\citep[see also][for a novel experiment using
the central entropies of X-ray emitting gas]{medezinski2016}. For typical galaxy
groups with $M_h{<}10^{13}\hmsol$ observed in the Sloan Digital Sky
Survey~\citep[SDSS;][]{york2000}, most halo finders cannot robustly separate
central galaxies from satellites, or identify which of two nearby groups hosts a
given satellite~\citep{campbell2015}. Therefore, despite the fact that halo
assembly bias is predicted to be stronger at lower mass, its observational
signature in SDSS groups~\citep{yang2007} has remained elusive to various
detection efforts~\citep{yang2006, wang2013, lacerna2014, lin16}.

Recently, \citet[][hereafter M16]{miyatake16} have discovered a strong halo
assembly bias effect among massive clusters~($\langle
M_{200m}\rangle{\simeq}1.9\times 10^{14}\hmsol$) using the \redmapper\ cluster
catalogue~\citep{rykoff14} derived from SDSS Data Release
8~\citep[DR8;][]{aihara2011}. They split the clusters into two subsamples based
on $\rmem$, the average projected distance of cluster membership candidates to
the central galaxy, expecting the two sets of clusters to have different average
halo concentrations. Using weak lensing, M16 discovered that the ratio between
the large-scale clustering biases of the two subsamples is
${\sim}1.64^{+0.31}_{-0.26}$, a $2.5\sigma$ deviation from unity. Using the same
cluster subsamples as M16, \citet{more16} measured their cross-correlations with
the SDSS photometric galaxy catalogue, and derived a much tighter constraint on
the bias ratio~($1.48\pm0.07$), which is $6.6\sigma$ above unity but still
consistent with M16. By adopting a linear model for the dependence of cluster
bias on $\rmem$, \citet{baxter16} found that a strong positive slope is required
to fit the angular clustering of \redmapper\ clusters split by $\rmem$,
confirming the results of M16 and \citet{more16}.

Intriguingly, the bias ratios measured by those studies with the M16 estimate of
$\rmem$ also exceed the level of halo assembly bias expected for similar
clusters in the $\lcdm$. In particular, cosmological $\lcdm$ simulations predict
that for a sample of massive haloes thresholded by the same comoving number
density as the \redmapper\ catalogue, the bias ratio between two subsamples
split by their dark matter concentrations is ${\simeq}1.24$, modulo minor
variations due to uncertainties in cosmological parameters. Observationally,
since $\rmem$ is usually estimated from the projected distances of relatively
bright satellites~($\sim{30}$ per cluster), it is likely a much cruder indicator
for assembly bias than the halo concentration measured in 3D from dark matter
particles in simulations, even if the underlying satellite galaxy concentration
somehow correlates better with halo assembly history than the dark matter
concentration. Therefore, we consider $1.24$ to be an upper limit of any
observable level of cluster assembly bias using $\rmem$, yet the M16 and
\citet{more16} measurements exceed this maximum value by $1.5\sigma$ and
$3.4\sigma$, respectively.

Without resorting to some new exotic physics, we are basically left with two
possible observational explanations. The first is that the bias ratio anomaly
could be merely a statistical fluke, which would disappear when a much larger
cluster sample becomes available~\citep{dalal2016}. On the other hand, there
could be some systematic uncertainties that are unaccounted for in the estimate
of $\rmem$, giving rise to a high bias ratio that is nonetheless irrelevant to
halo assembly bias. In this paper, we examine the potential systematic
uncertainties associated with the estimate of $\rmem$, particularly the impact
of projection effects due to having multiple clusters on the same line-of-sight,
and re-analyze the halo assembly bias signal within \redmapper\ using a new
$\rmem$ estimator that is robust to such projections.

We organize the paper as follows. In \S~\ref{sec:pmem} we investigate the
possible imprint of projection effects on the distribution of membership
probabilities in \redmapper. We present a null test diagnostic for projection
effects and develop a new $\rmem$ estimator that passes this test in
\S~\ref{sec:nulltest}. In \S~\ref{sec:revisit} we predict the level of
observable assembly bias in \redmapper\ after accounting for the scatter between
$\rmem$ and halo concentration using mock cluster catalogues, and compare the
prediction to the signal measured in \redmapper\ from weak lensing. Finally, we
summarize our findings and discuss their implications for future surveys in
\S~\ref{sec:conc}.

\section{Projection Effect on Membership Probabilities}
\label{sec:pmem}

\begin{figure*}
\begin{center}
    \includegraphics[width=0.9\textwidth]{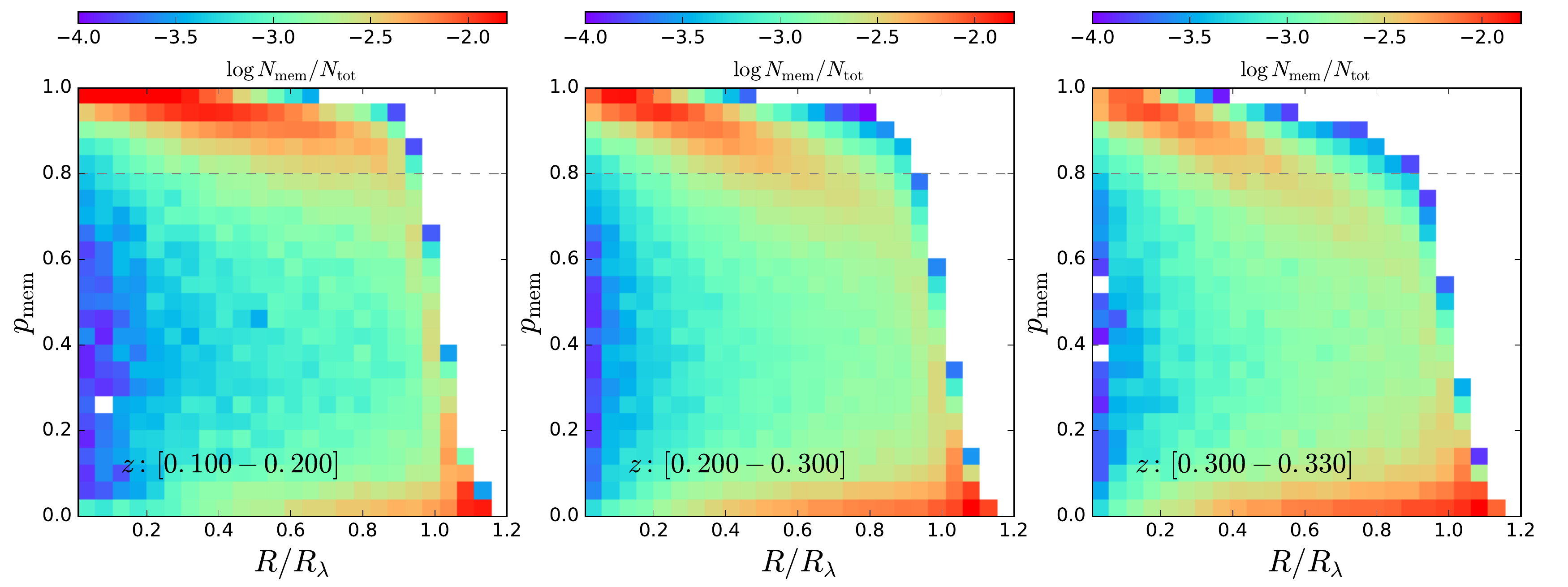} \caption[]{Member
    galaxy number distributions on the membership probability vs.\ projected
    distance~(normalized by $\rlam$ defined by Equation~\ref{eqn:rlam}) plane,
	for clusters within $z=[0.10, 0.20]$~(left), $[0.20, 0.30]$~(middle),
	and $[0.30, 0.33]$, respectively. In each panel, the 2D distribution of
	logarithmic galaxy numbers~(normalized by total number of galaxies) is
	color-coded by the color bar on the top. The low-$\pmem$ galaxies become
	more abundant and reach lower $\rmem$ with increasing redshift. The
	horizontal line indicates the minimum $p_m=0.8$ that we used to define
	$\rzu$.}
    \label{fig:pmemmap}
\end{center}
\end{figure*}

Due to the lack of accurate distances, photometric cluster-finders suffer from
various types of projection effects. In this paper, ``projection effect'' refers
to the contamination of cluster membership probabilities by other systems along
the same line-of-sight but outside the virial radius of that cluster. We
emphasize that this particular effect is different from the common perception of
projection in detecting clusters, i.e., the blending of multiple systems along
the same line-of-sight into one large cluster~\citep{erickson2011, noh2012}. In
the case of \redmapper~\citep{rykoff14}, there are two steps designed to remove
the impact of projection effects on the estimate of the cluster richness
$\lambda$~\citep[for an alternative scheme see][]{castignani2016}:
\begin{itemize}
    \item The member galaxy candidates of each cluster are searched within some
    finite aperture with a physical radius $\rlam$ that depends on the richness of the
    cluster $\lambda$,
    \begin{equation}
        \rlam \simeq (\lambda/100)^{0.2}\,\mathrm{Mpc}/h,
        \label{eqn:rlam}
    \end{equation}
    so that all the member candidates are found within
    $\rlam$ of the cluster center and are
    assigned a membership probability $p_{\mathrm{mem}}$.
    \item A percolation procedure was serially applied to all the clusters,
    assigning each candidate member galaxy a probability $p_{\mathrm{free}}$ to
    account for the possibility of it being already assigned to another cluster
    along the same line-of-sight.
\end{itemize}
In addition, there are two other probabilities that describe the soft cuts in
radius~($\Theta_R$) and magnitude~($\Theta_I$). Therefore, each galaxy within
the aperture will have an effective membership probability
$p_m{\equiv}p_{\mathrm{mem}}\times p_{\mathrm{free}}\times\Theta_R \times
\Theta_I$. Note that although M16 used $p_{\mathrm{mem}}\times
p_{\mathrm{free}}$ as the effective membership probability when estimating
$\rmiya$, their large vs.\ small-$\rmiya$ split is unchanged when $p_m$ is
adopted. The galaxy surface number density profile of a cluster can then be
estimated as
\begin{equation}
    \Sigma_g(R) = \frac{1}{ 2 \pi R \Delta R} \sum_{j} p^j_m\;\;\;\;\mbox{for}\;R_{j}\in R\pm\Delta R/2,
    \label{eqn:defsigg}
\end{equation}
and the cluster richness is
\begin{equation}
    \lambda = \sum_i p^{i}_m = 2 \pi \int_0^{R_0} R\, \Sigma_g(R)\, \dd R, \label{eqn:deflambda}
\end{equation}
where the index $i$ runs over all the membership galaxy candidates within the aperture.

Meanwhile, to compute $\avg{R_{\mathrm{mem}}}$, M16 applied the same membership
weights to the projected distances
\begin{equation}
    \rmiya = \frac{\sum_i (p^{i}_{m} \, R_i)}{\lambda} = \frac{2 \pi}{\lambda} \int_0^{R_0} R^2\, \Sigma_g(R)\, \dd R.
    \label{eqn:defrmiya}
\end{equation}
In the case that the $p^{i}_m$ values are unbiased, $\rmiya$ is the correct
estimator for $\rmem$. In cluster finders, the membership probability $p_m$ is
closely tied to the {\it expected} density contrast between cluster members and
background galaxies. In particular, the current version of the \redmapper\
algorithm models the galaxy distribution surrounding the center of each cluster
as the sum of an intrinsic cluster member component and a uniform background
component, and then derives membership probabilities that are consistent with
this two-component model~\citep{rykoff14}. However, the background galaxy
portion of the model is determined globally across the whole survey, and the
$p_m$ assignment may thus be affected by fluctuations in the local density of
background galaxies around individual clusters.

For instance, if the cluster were observed in a very crowded area on the sky, it
is plausible that the cluster finder would incorrectly assign very low but
non-zero values of $p_m$ to some background galaxies at large $R$, where the
intrinsic galaxy number density profile $\Sigma_g(R)$ begins to drop
precipitously. As a result, the observed $\Sigma_g(R)$ is more extended and
flattened at large $R$, affecting the estimation of both $\lambda$ and $\rmiya$
simultaneously. To examine the impact of projection effects on the estimate of
$\lambda$ in \redmapper, \citet{rykoff14} performed a Monte Carlo test by
randomly injecting simulated clusters with known $\lambda_{\mathrm{true}}$ onto
the actual observed background galaxy map, and measured the distribution of
$\lambda_{\mathrm{obs}}$ returned by \redmapper\ at fixed
$\lambda_{\mathrm{true}}$. They found that most clusters fall within a tight
locus around $\lambda_{\mathrm{obs}}{\simeq}\lambda_{\mathrm{true}}$, suggesting
little systematic bias in the estimate of richness due to this global background
model. However, comparing Equations~\ref{eqn:deflambda} and~\ref{eqn:defrmiya},
we can immediately tell that the estimation of $\rmiya$ is much more sensitive
to the shape of $\Sigma_g(R)$ at large $R$ than that of $\lambda$. Therefore,
even at fixed (or minimally biased) $\lambda$, the average membership distance
estimated from Equation~\ref{eqn:defrmiya} could be systematically biased to
larger values of $\rmiya$ in crowded areas than in isolated ones. Note that the
contamination should persist in high-$p_m$ galaxies at some reduced level, but
its impact on $\rmiya$ is likely much smaller.

\begin{figure*}
\begin{center}
    \includegraphics[width=0.9\textwidth]{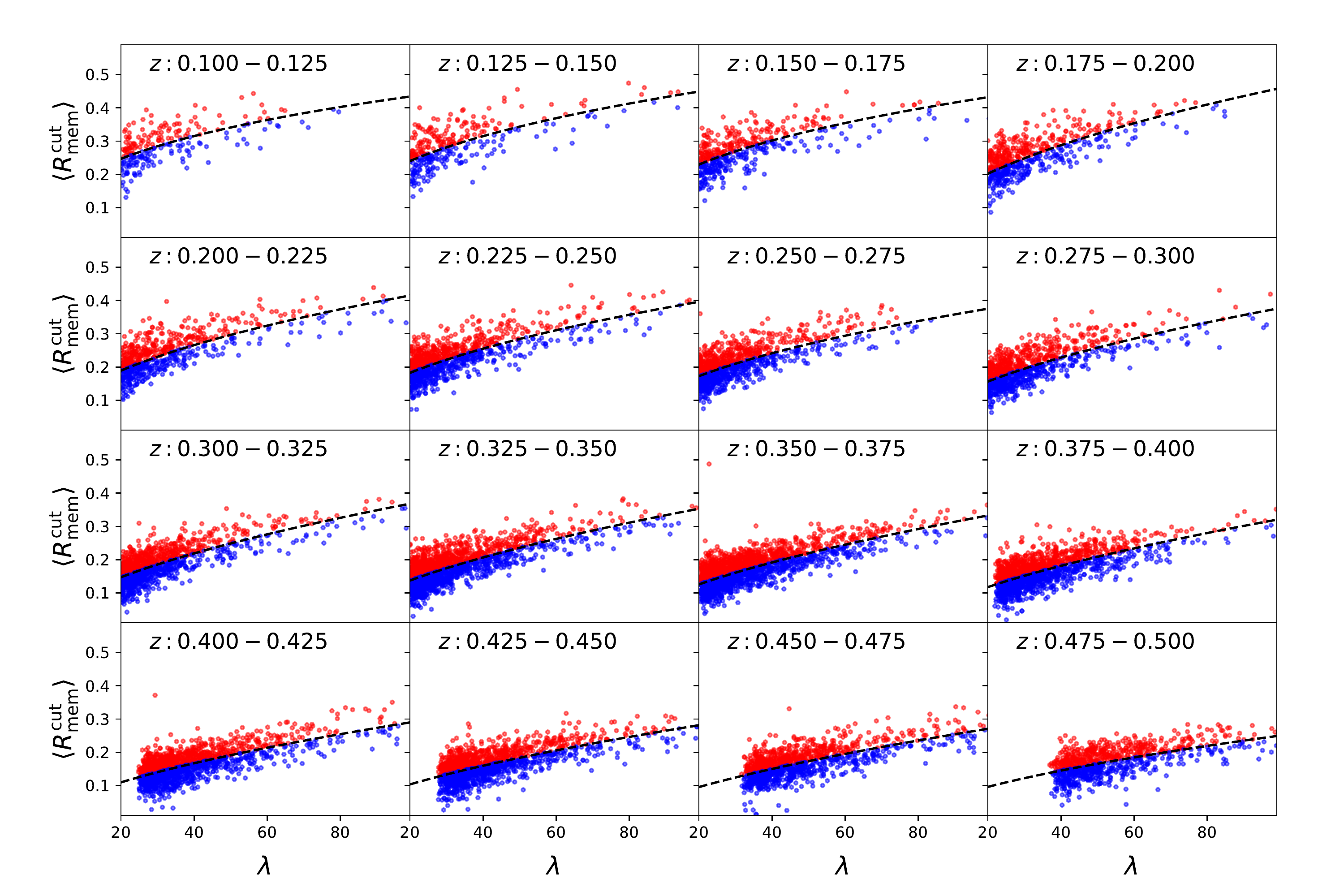}
    \caption[]{Subsamples of large~(red) and
	small~(blue) $\rzu$ clusters on the $\rzu$ vs.\ $\lambda$ plane at 16
	different redshifts, using $p^{\mathrm{crit}}_{\mathrm{mem}}=0.80$. The
	dashed curve in each panel indicates the median $\rzu$ as
    a function of richness at that redshift. } \label{fig:rmemsplit}
\end{center}
\end{figure*}

Fig.~\ref{fig:pmemmap} shows the member galaxy number distributions on the 2D
plane of $p_m$ vs.\ $R/\rlam$, for clusters in redshift ranges $[0.1,
0.2]$~(left), $[0.2, 0.3]$~(middle), and $[0.3, 0.33]$~(right), respectively. As
expected, the average $p_m$ is a declining function of $R/\rlam$. More
importantly, the distribution of $p_m$ at fixed $R/\rlam<0.9$ is bimodal,
consisting of one population with extremely low values of $p_m$ and another with
$p_m{>}0.6$. At large distances, the low-$p_m$ galaxies becomes dominant over
the high-$p_m$ ones. This scale dependence of the relative fraction of the two
populations should vary from cluster to cluster, depending on the level of
contamination from local background.  The low-$p_m$ population also becomes
progressively more prominent with increasing redshift, while reaching further
into the inner part of clusters. Therefore, placing a simple cut on $R/\rlam$
would not eliminate the low-$p_m$ galaxies at high redshifts. As described
earlier, if the low $p_m$ values are more often incorrectly assigned in crowded
regions on the sky, $\rmiya$ estimated from Equation~\ref{eqn:defrmiya} would be
biased high for clusters in those regions.

There are two avenues for addressing the potential background contamination
issue in membership assignments.  One would be to improve the \redmapper\
algorithm to account for the locally-varying background in determining $p_m$, in
which case the $\rmiya$ estimator may be usable again. The alternative is to
update the way we estimate $\rmem$. In this work we adopt the latter route as it
can be trivially explored with the existing public \redmapper\ catalogue.

In order to eliminate any problems caused by the contaminated low-$p_m$
galaxies, we define a variant of the average membership distance by placing a
cut on $p_m$,
\begin{equation}
    \rzu = \frac{\sum_i (p^{i}_{m} \, R_i)}{\sum_i
        p^{i}_m}\;\;\;\;\mbox{for}\;p^{i}_m>p^{\mathrm{crit}}_{\mathrm{mem}},
    \label{eqn:defrzu}
\end{equation}
assuming the fractional amount of contamination decreases with increasing $p_m$. The choice of
$p^{\mathrm{crit}}_{\mathrm{mem}}$ must satisfy two criteria:
\begin{itemize}
    \item The projection effect is eliminated as determined using a suitable null test discussed below.
    \item The two cluster subsamples split by $\rzu$ should exhibit different concentrations of their galaxy
    surface number density profiles.
\end{itemize}
After running both empirical tests~(discussed further below) over a grid of
$p_m$ cuts, we pick the value of $p^{\mathrm{crit}}_{\mathrm{mem}}=0.8$, as
shown by the dashed horizontal lines in Fig.~\ref{fig:pmemmap}.  Under this cut,
the effective richness, calculated by summing all the $p_m$ above
$p^{\mathrm{crit}}_{\mathrm{mem}}$, shrinks significantly at high redshifts,
introducing some extra scatter into the ranking-order of $\rmem$ at fixed $z$.
However, even at the highest redshift bin~($z{\sim}0.30$)
almost all~(${>}98\%$) the clusters still retain more than $6$ member galaxy candidates
for computing $\rzu$, and the fraction~(weighted by $p_m$) of all member
galaxies with $p_m>0.8$ is $\sim{0.55}$.

Fig.~\ref{fig:rmemsplit} illustrates the division of clusters into large and
small-$\rzu$ subsamples at 16 different redshifts bins from $z=0.1$ to $0.5$. At
each redshift $z$, we separate the clusters by the median of their $\rzu$ as a
function of $\lambda$~(black dashed curve).  Note that beyond $z=0.33$ the
sample systematically misses low-richness clusters, as the luminosity threshold
for membership galaxies~($0.2L_*$) hits the magnitude limit of the DR8
catalogue~($i{<}21$). We nonetheless make use of all clusters up to $z=0.50$ for
the null test of projection effects~(discussed below in \S~\ref{sec:nulltest}),
which does not require sample-completeness in richness. However, we will limit
our assembly bias analysis~(see \S~\ref{sec:revisit}) to clusters between $0.1$
and $0.33$, where the \redmapper\ catalogue is approximately volume-complete.

\begin{figure}
\begin{center}
    \includegraphics[width=0.5\textwidth]{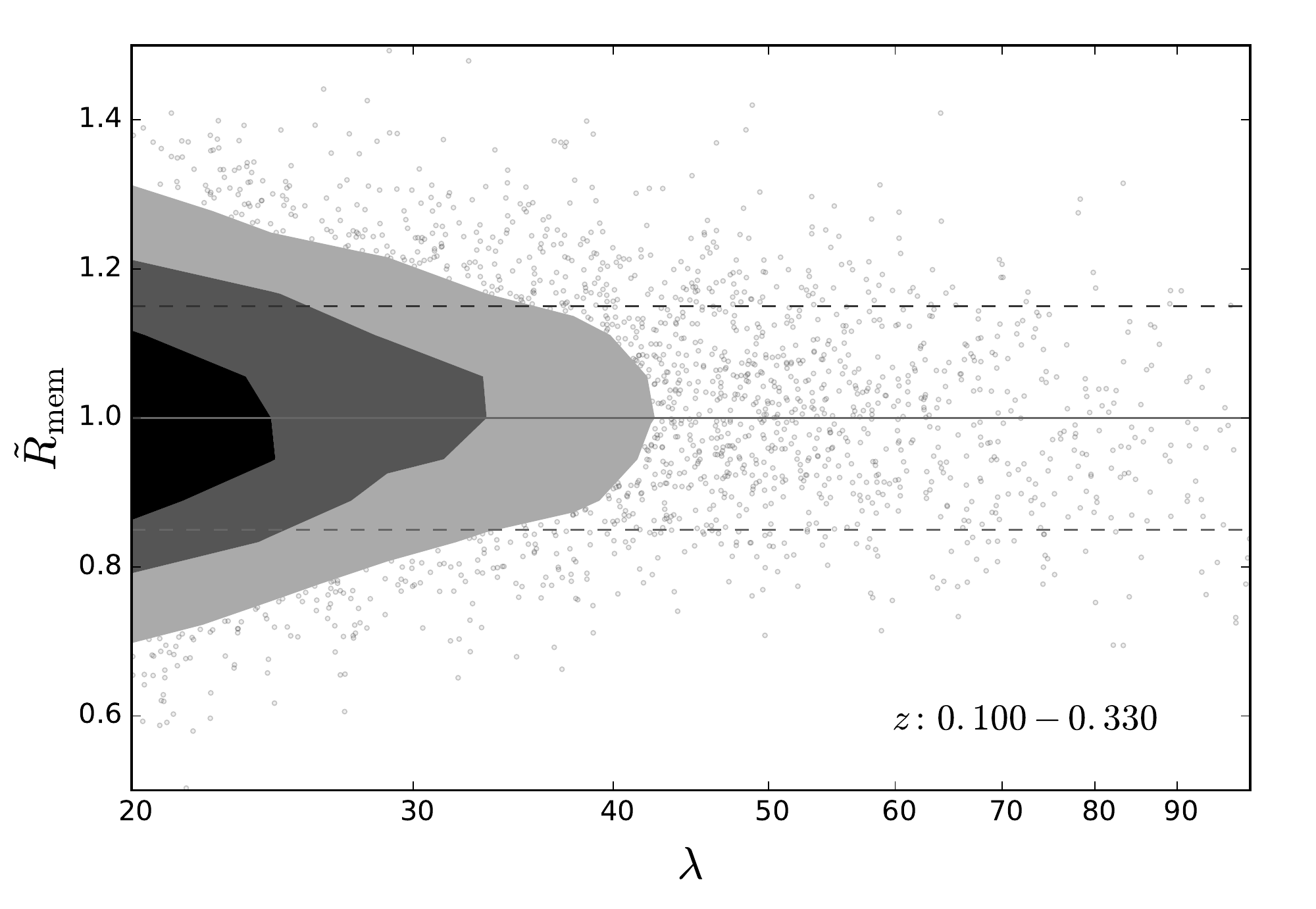} \caption[]{Distribution of the reduced average
    membership distance $\smem$ measured as a function of cluster richness $\lambda$. The three contour lines
    enclose $25\%$, $50\%$, and $75\%$ of the clusters, respectively.
Unlike the distributions of $\rzu$ shown in Figure.~\ref{fig:rmemsplit}, the mean~($1$)
    and the standard deviation~($0.15$) of the distributions of $\smem$ are uniform across the
   entire redshift range of the \redmapper\ sample.}
    \label{fig:smemres}
\end{center}
\end{figure}

Fig.~\ref{fig:rmemsplit} also shows that both the average amplitude and the
scaling of median $\rzu$ with $\lambda$ vary smoothly with redshift. In order to
quantify the deviations of individual $\rzu$ from the median and compare them
across different $\lambda$ and $z$, we fit a power-law relation
$\overline{\rzu}(\lambda, z)=a(z)\lambda^{b(z)}$ to the median relation at each
$z$, and spline-interpolate the best-fitting parameters $a(z)$ and $b(z)$ over
the entire redshift range. In this way, we can classify a cluster with given
$\rzu$ into large or small-$\rzu$ subsamples directly based on its $\lambda$ and
$z$.

Furthermore, we normalize the deviations of $\rzu$ from
$\overline{\rzu}(\lambda, z)$ by the standard deviation of $\rzu$ at fixed
$\lambda$ and $z$, i.e., defining a {\it reduced} average membership distance
$\smem{=}1 + 0.15\left(\rzu - \overline{\rzu}(\lambda,
    z)\right)/\sigma_{\rzu}(\lambda,z)$, so that the distribution of $\smem$ always has a mean of unity and a
dispersion of $0.15$~(arbitrarily chosen). As a result, the probability distribution
functions of $\smem$ are almost identical across any fixed richness and
redshift.  Fig.~\ref{fig:smemres} shows the cluster number distribution on the
$\smem$ vs.\ $\lambda$ plane, with two dashed horizontal lines indicating the
1-$\sigma$ range of $\smem$ at fixed $\lambda$.  To facilitate the comparison
between the M16 results and ours, we apply the same transformation to the
distribution of $\rmiya$ to obtain $\smiya{=}1 + 0.15\left(\rmiya -
\overline{\rmiya}(\lambda, z)\right)/\sigma_{\rmiya}(\lambda,z)$.  By
construction, the distribution of $\rmiya$~(not shown here) is similar to
Fig.~\ref{fig:smemres}.

\begin{figure}
\begin{center}
    \includegraphics[width=0.5\textwidth]{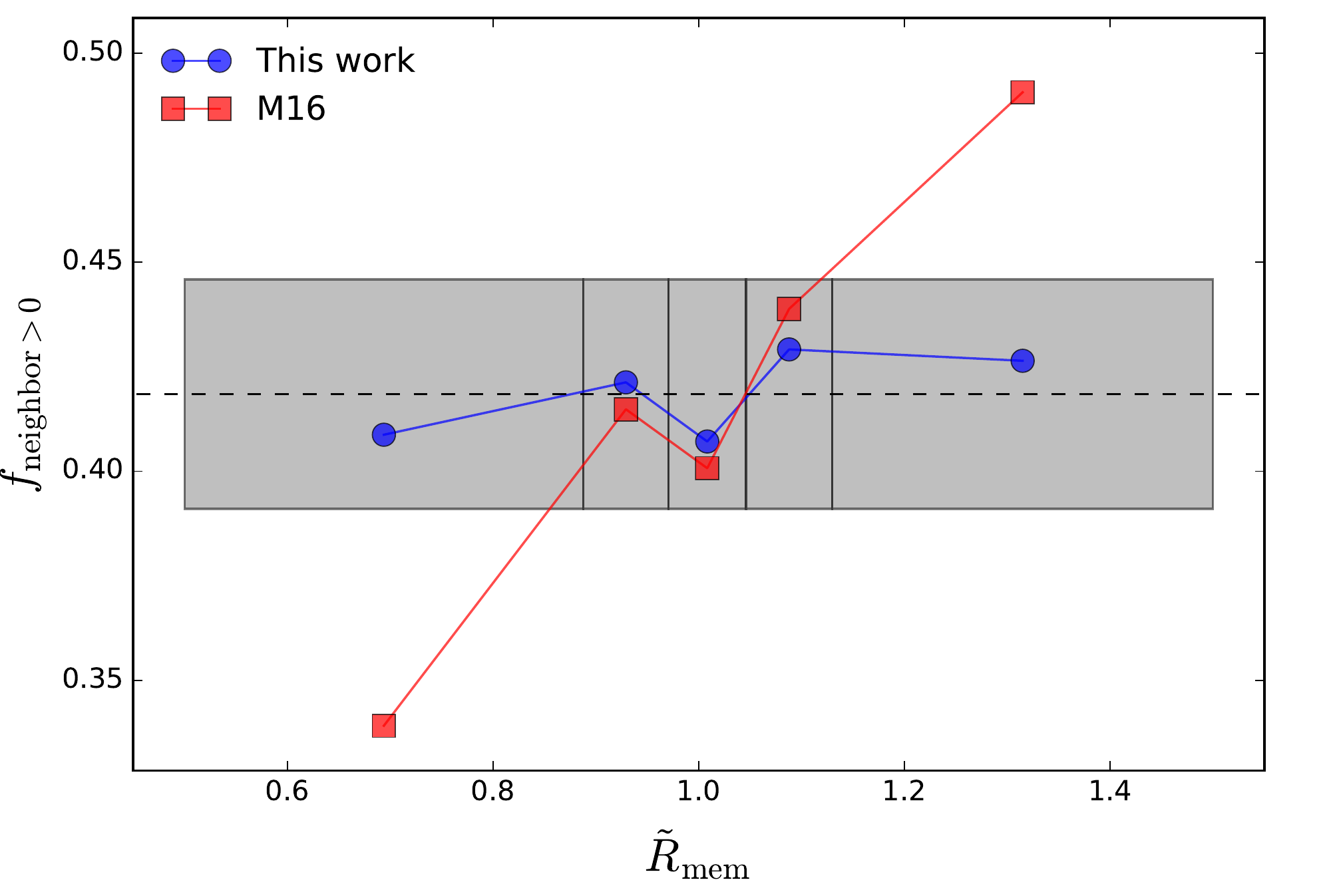} \caption[]{
        Fraction of clusters that have one or multiple physically unassociated clusters along the same
	line-of-sight as a function of $\smem$~(blue circles) or $\smiya$~(red squares). For a given cluster
	at $z\in[0.3, 0.33]$, we count any other clusters that are observed at $\Delta z{=}0.02$ away and
    within a solid angle that corresponds to $3.2\hmpc$ at $z$ as ``physically unassociated neighbors''. The
    gray band indicates the $1\sigma$ uncertainty for a $\rmem$ estimator free of projection
    effects.
}
    \label{fig:fnnb}
\end{center}
\end{figure}

Fig.~\ref{fig:fnnb} provides a direct evidence of the impact of projection
effects in the estimate of $\rmiya$. For a given cluster at $z\in
[0.30,0.33]$,\footnote{We pick this redshift bin for Fig.~\ref{fig:fnnb} and
later the mark correlation functions because it shows the most prominent
population of low-$p_m$ galaxies, and therefore is the most susceptible to
projection effects.} we identify all the physically unassociated~(explained
further below) clusters that happen to lie along the same line-of-sight, and
compute the fraction of cluster sight-lines that have one or multiple such {\it
apparent} neighbors as a function of $\smiya$~(red squares) or $\smem$~(blue
circles). Specifically, we search secondary clusters within a solid angle that
corresponds to $d=3.2\hmpc$ at the redshift of the primary, and at a redshift at
least $\Delta z=0.02$ away from the primary. This value of $\Delta z$ is $33\%$
larger than the cluster photo-z uncertainty~\citep[$\sigma_z=0.015$ at
$z=0.3$;][]{rykoff14}, so that the two systems are unlikely to be physically
associated. We have also verified that using different choices of $d\in[2, 4]$
and $\Delta z\in[0.03, 0.04]$ do not qualitatively change the result of this
experiment. The gray band indicates the $1\sigma$ uncertainty for a $\rmem$
measure that is free of projection effects, derived from a Monte Carlo test in
which we re-measure $f_{\mathrm{neighbor}>0}$ after randomly shuffling the
$\smiya$ values among clusters and repeat the exercise $50000$ times. The
uncertainties are identical across all five $\smiya$ bins as there are equal
numbers of clusters in each bin by design.

Clearly, if a cluster is detected with some other unassociated systems projected
close to its sight-line, it is significantly more likely to have a higher
estimated value of $\rmiya$ than one without. In contrast, our new $\rzu$
estimator is consistent with being free of such projection effects, showing a
uniform $f_{\mathrm{neighbor}>0}$ as a function of $\smem$. Therefore,
Fig.~\ref{fig:fnnb} confirms our expectation that the background contamination
in low-$p_m$ galaxies causes systematic biases in the estimate of $\rmiya$, and
by simply removing them when defining $\rzu$ we can significantly reduce or even
eliminate the impact of projection effects on $\rmem$.

However, one drawback of the $f_{\mathrm{neighbor}}$ test in Fig.~\ref{fig:fnnb}
is that, it probes only one distance scale $d$ each time, while the halo
assembly bias exists on all scales above 1-halo~\citep{sunayama2016} and is
usually measured on scales above $10\hmpc$ --- a non-parametric null test that
can probe all scales is required.

\section{A Null Test for Diagnosing Projection Effects}
\label{sec:nulltest}

We employ the angular mark cross-correlation function between two cluster
samples that are well-separated in redshift as a null test for diagnosing
projection effects in each $\rmem$ definition.  The angular mark correlation is
defined as~\citep{stoyan94, beisbart2000, sheth2005, skibba06, harker2006}
\begin{equation}
    M_{R_{\mathrm{mem}}}(\theta) = \frac{1+\mathcal{W}(\theta)}{1 + w(\theta)},
    \label{eqn:markcorr}
\end{equation}
where $\mathcal{W}$ and $w$ are the $\rmem$-weighted and regular angular
correlation functions, respectively, and $\theta$ is the angular distance on the
sky. Within each cluster sample the marks $\smiya$ and $\smem$ are normalized to
have a mean of unity. Therefore, if the marks of a pair of clusters from two
different samples separated by distance $\theta$ do not correlate with each
other, the mark correlation $M$ would be exactly unity at $\theta$. In practice
the angular mark correlation function can be directly computed via dividing the
mark-weighted pair counts $WW$ by the unweighted ones $DD$
\begin{equation}
    M_{R_{\mathrm{mem}}}(\theta) = \frac{WW(\theta)}{DD(\theta)},
    \label{eqn:markcorr2}
\end{equation}
thus avoiding the need of random catalogues. Compared to the
$f_{\mathrm{neighbor}}$ test in Fig.~\ref{fig:fnnb},
$M_{R_{\mathrm{mem}}}(\theta)$ is an indirect but statistically more powerful
null test of projection effects, as it probes all angular scales at once by
using all cluster pairs in the catalogue.

\begin{figure}
\begin{center}
    \includegraphics[width=0.5\textwidth]{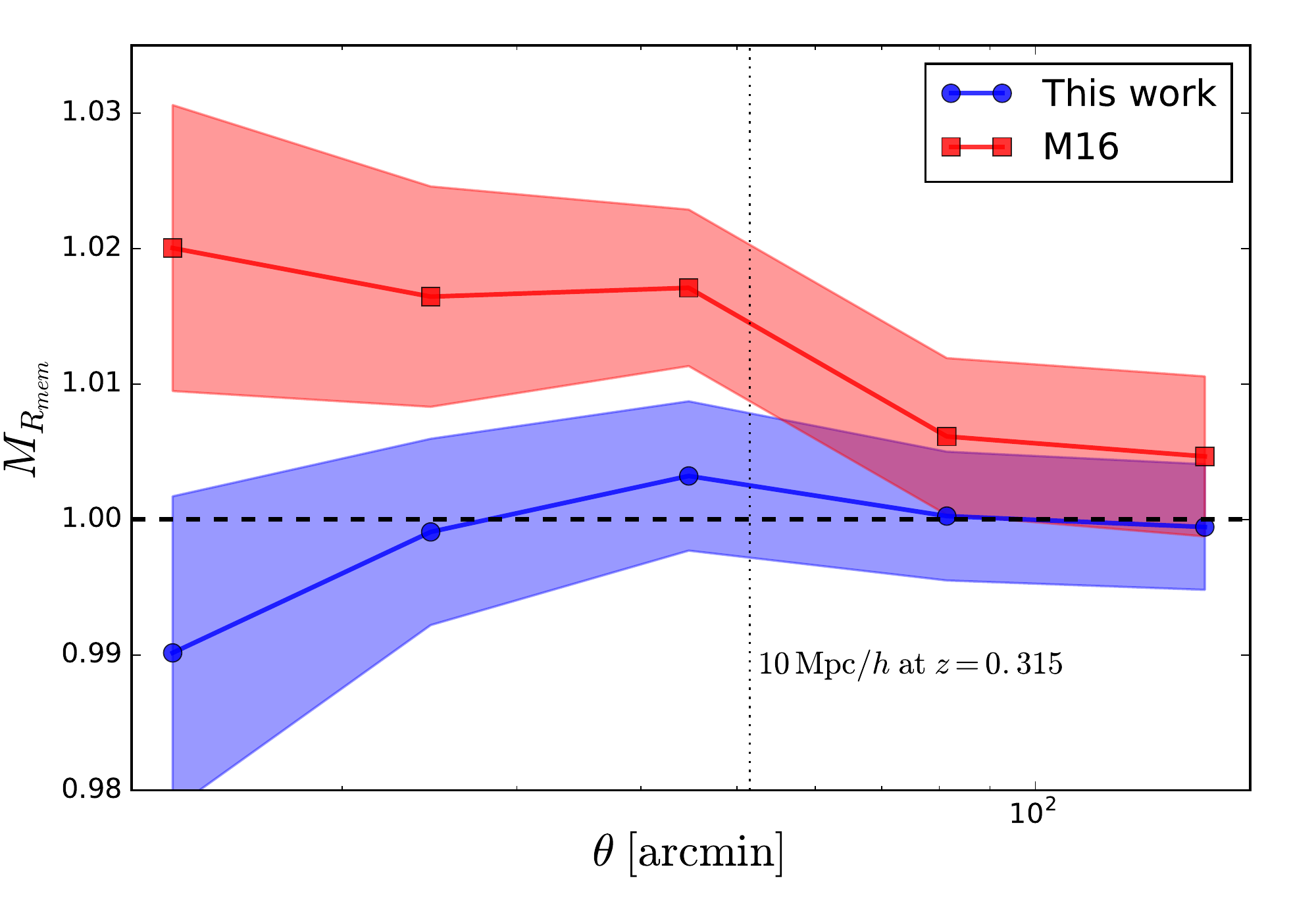} \caption[]{Angular mark cross-correlation
    functions between clusters at $z\in [0.30, 0.33]$ and the joint cluster
	samples at $z\in[0.10, 0.28]$ and $z\in[0.35, 0.50]$, using $\smiya$~(red) and $\smem$~(blue) as
    marks, respectively. The vertical dashed line indicates the angular distance that corresponds to a
    physical separation of $10\,\hmpc$ at $z=0.3$.} \label{fig:mktheta}
\end{center}
\end{figure}
Fig.~\ref{fig:mktheta} shows the angular mark correlation functions between clusters with $z \in [0.30, 0.33]$
and all other clusters with either $z\in [0.10, 0.28]$ or $z\in [0.35, 0.50]$ in \redmapper, using
$\smiya$~(red) and $\smem$~(blue) as markers, respectively. Using a large-volume $N$-body
simulation~($2.5^3\cubichgpc$; described further below), we have verified that the redshift separation between
the slices, $\Delta z=0.02$, which corresponds to a comoving distance of $52\hmpc$ at $z=0.3$, is large enough
so that the mark correlation function~(using halo concentrations as markers) is consistent with unity, i.e.,
there is little physical assembly bias left.

This angular marked statistics $M_{\rmem}$ serves as our null test for the
presence of projection effects in any estimators of $\rmem$: for a $\rmem$
estimator free of membership contaminations, its $M_{\rmem}$ should be unity on
all angular scales because two cluster samples in the same sky area but
different redshift ranges are not physically associated. Clearly, our new
estimator $\smem$ passes this null test on all measured scales $\in[10,
200]$~arcmin, while the M16 estimator $\rmiya$ does not, showing spatially
coherent and statistically significant correlation between pairs of physically
unassociated clusters on the sky.

Combining Figs.~\ref{fig:fnnb} and~\ref{fig:mktheta}, we establish that the
$\rmiya$ selection is affected by projection effects that have contaminated the
photometric galaxy membership probabilities at low values of $p_m$, while the
$\rzu$ selection is almost immune to such projection effect, by separating high
and low-concentration clusters based only on member galaxies with high $p_m$.

\begin{figure}
\begin{center}
    \includegraphics[width=0.5\textwidth]{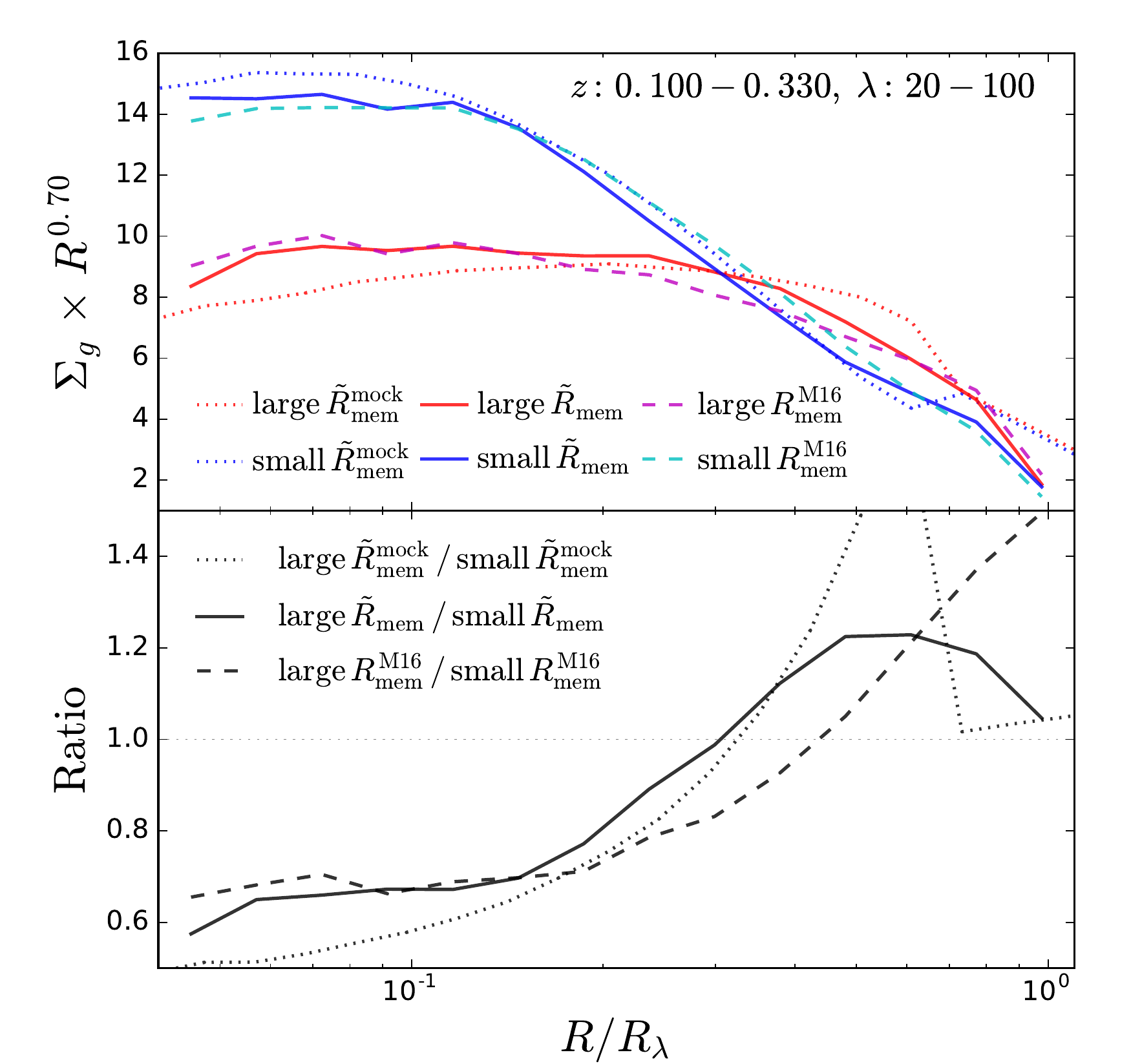} \caption[]{Member galaxy surface density profiles
	$\Sigma_g$~(with $p_m$-weighting) of the large~(red or magenta) and small~(blue or cyan)-$\rmem$
	clusters, split by $\smiya$~(dashed) and $\smem$~(solid), respectively. The dotted curves are the
	surface density profiles predicted by a mock cluster catalogue that mimics the $\rzu$-split.  The
	bottom panel shows the ratio between the $\Sigma_g(R)$ of high and low-concentration clusters within
	each definition of $\rmem$.
    % The split in galaxy concentration by $\smem$ relies on the inner galaxy
	% density profiles, whereas $\rmiya$ is more affected by the large number of low-$\rmiya$ galaxies near
	% the aperture edge.
    The mock cluster catalogue correctly reproduces the contrast between the large
	and small-$\rzu$ clusters~(compare solid to dotted curves), especially the distance at which the ratio
	crosses unity in the bottom panel.} \label{fig:dsgal}
\end{center}
\end{figure}

One might worry that the subsamples split by $\rzu$, which uses only ${\sim}55\%$ of the
total membership candidates, do not have average mass/galaxy density profiles
with distinct concentrations. To find out, we compare the galaxy surface number
density profiles of large and small-$\rmem$ clusters in Fig.~\ref{fig:dsgal},
split by $\rmiya$~(dashed) and $\rzu$~(solid), separately. The $\Sigma_g$
profiles are computed using Equation~\ref{eqn:defsigg} with the $p_m$-weighting
on each galaxy. Also note that when computing the $\Sigma_g$ profiles for
$\smem$-selected clusters, we do not place any cut on $p_m$ and employ all
member candidates with $p_m>0$. Both ways of splitting samples return two sets
of high and low-concentration $\Sigma_g$ profiles, with similar amplitudes at
small $R$.

The dashed curves in Fig.~\ref{fig:dsgal} are the measurements from a mock
cluster catalogue and the $R_{\mathrm{mem}}^{\mathrm{mock}}$ estimator is
designed to mimic the behavior of $\rzu$ in \redmapper~(see
\S~\ref{sec:revisit}). We will describe the detail of this mock cluster
catalogue and the definition of $R_{\mathrm{mem}}^{\mathrm{mock}}$ in the
section below, and then come back to Fig.~\ref{fig:dsgal}.

\section{A Re-analysis of Cluster Assembly Bias}
\label{sec:revisit}

As alluded to in \S~\ref{sec:intro}, the cluster assembly bias signal revealed
by $\rmem$ in observations would likely be diluted compared to the signal
predicted from simulations using dark matter concentration~($c$), due to the
large scatter of $\rmem$ at fixed $c$. In the case of $\rzu$, the scatter
$\sigma_{\rzu|c}$ for any given cluster mass has three contributing sources:
\begin{itemize}
    \item[1)] The intrinsic spread in the distribution of the underlying galaxy concentration $c_g$ at fixed $c$.
    \item[2)] The observational scatter between $\rzu$ and $c_g$ within clusters, due to Poisson fluctuations
    in the galaxy number density profile, the use of projected distances~(instead of radial distances), and
    the small aperture size for selecting members.
    \item[3)] The systematic uncertainty in $\rzu$ due to some residual contamination in the membership
    probabilities of the $p_m{>}0.8$ galaxies.
\end{itemize}
Among the above three, we expect the observational scatter between $\rzu$ and
$c_g$ to be the dominant source of signal dilution. For the spread in the
distribution of $c_g$ at fixed $c$, it depends on how well the satellite
galaxies follow the orbits, and hence the spatial distribution, of dark matter
particles. Although the radial distribution of subhalos at or above some fixed
{\it present-day} mass $m_{0}$ is flatter than the density profile of the host
halo~\citep{diemand2004, gao2004}, the radial profile of subhalos selected on
their tidally truncated mass~(or the peak mass over their assembly history
$m_{\mathrm{peak}}$) traces the dark matter very well, except at the inner
radius where dynamical friction would affect the distribution of more massive
subhalos~\citep{nagai2005, han2016, vdbosch2016}. Since we expect galaxies
selected on stellar mass or luminosity correspond closely to subhalos selected
on $m_{\mathrm{peak}}$ rather than $m_{0}$, it is reasonable to expect the
intrinsic scatter in $c_g$ at fixed $c$ to be subdominant compared to the
observational scatter in $\rzu$ at fixed $c_g$. For the residual systematic
uncertainties in $\rzu$, we expect it to be negligible at the noise level of
\redmapper, based on the null test in Figure~\ref{fig:mktheta}. The mark
correlation has a slight dip below unity in the lowest angular distance bin at
$15\,\mathrm{arcmin}$, but it is more likely a statistical fluctuation within
the uncertainty and should not be caused by projection effects, which only leads
to correlations above unity.

Therefore, before measuring the assembly bias signal using $\rzu$ in \redmapper,
we need to carefully examine the impact of the scatter between $\rzu$ and $c_g$
on the observable level of signal, using mock catalogues of massive clusters and
red-sequence galaxies. For the mock cluster catalogue, we make use of the
\rockstar~\citep{behroozi13} halo catalogue derived from the $z=0.25$ output of
the \bigmdpl\
simulation\footnote{\url{https://www.cosmosim.org/cms/simulations/bigmdpl}}. The
simulation was evolved under Planck cosmology~\citep{planck15} with $3840^3$
particles in a cubic box of $2.5\hgpc$ on a side~\citep{klypin16}.  The large
box size is necessary to ensure that the large-scale bias measurements are not
limited by cosmic variance. We describe the construction of mock cluster and
red-sequence galaxy catalogues in detail below.

We employ the \ihod\ framework recently developed by \citet{zu15, zu16, zu17} to populate halos with mock
galaxies that reproduce the low-redshift lensing and clustering measurements of SDSS galaxies, and from them
we further select a red-sequence population based on the halo-quenching model of \citet{zu16}. Similarly, the
stellar mass and color distributions of the mock galaxies closely mimic the observed galaxies at
$z=0.25$~\citep{zu17}, because the best-fitting parameters for the \ihod\ halo-quenching prescription are
derived from the spatial clustering and galaxy-galaxy lensing measurements of SDSS galaxies at $z\in[0.0,
0.3]$.

Within each halo, most of the subhalos are not well-resolved in \bigmdpl\ due to
its relatively low mass resolution~($m_p=2.34\times10^{10}\hmsol$). Therefore,
instead of using the positions of $m_{\mathrm{peak}}$-selected subhalos, we
assign 3D distances to satellite galaxies based on an NFW profile with galaxy
concentration $c_g{=}0.86 c$. The slightly under-concentrated galaxy
distribution is preferred by the small-scale clustering and g-g lensing
measurements of SDSS galaxies~\citep{zu15} and the observed radial distribution
of satellites inside
clusters~\citep{yang2005,chen2006,budzynski2012,wang2014,zenteno2016}.

The richness of each mock cluster is assigned by counting the number of red
galaxies within $r_{200m}$ and above the stellar mass threshold
$M^{\mathrm{lim}}_*=1.66\times10^{10}\hhmsol$. We pick this value of
$M^{\mathrm{lim}}$ so that the mean halo mass of our mock clusters with
$\lambda$ between $20$ and $100$ is the same as in the \redmapper\
catalogue~\citep[$\langle M_h\rangle{\simeq}1.86\times 10^{14}\hmsol$;
see][]{miyatake16, simet16}. Note that the richness defined in our mock uses a
different aperture than in \redmapper, where the richness is derived using the
richness-dependent aperture $\rlam$ via a complicated iterative scheme. This
difference, however, does not affect the comparison between the mock and
\redmapper\ cluster samples, as the mock richness is merely used for selecting
clusters, and we are not comparing the two on individual cluster basis.

Finally, we need to incorporate the small aperture effect in the $\rzu$
estimator due to the $p_m{>}0.8$ selection~(see Fig.~\ref{fig:pmemmap}).
Ideally, we would want to simulate the $p_m$ values first by running the
\redmapper\ cluster finder over a mock SDSS imaging catalogue, and select only
high-$p_m$ mock galaxies when deriving $\rzu$ for the detected mock clusters.
However, this is a rather difficult task, requiring a near-perfect understanding
of the photometric redshift properties of SDSS galaxies up to $z{=}0.55$. In
reality, since the amount of background contamination in $\rzu$ is negligible,
the effective aperture size associated with the $p_m{>}0.8$ selection should not
depend on the local background. Therefore, we can bypass the task of simulating
realistic $p_m$, and assume the small aperture effect due to a cut in $p_m$ can
be roughly mimicked by adopting an effective aperture size that is some fixed
fraction $a$ of $\rlam$ in the mock. We empirically determine the value of $a$
to be $0.65$, by enforcing that the fraction of $R{<}\rlam$ galaxies included by
this $R{<}a\rlam$ cut is $0.55$, the same as that included by the $p_m>0.8$ cut
in \redmapper. For each mock cluster, we now define $\langle
R^{\mathrm{mock}}_{\mathrm{mem}}\rangle$ as the average 2D membership distance
of galaxies within $0.65\rlam$.

To check whether the scatter between $\langle
R^{\mathrm{mock}}_{\mathrm{mem}}\rangle$ and $c_g$ is comparable to the
intrinsic scatter between $\rzu$ and $c_g$ within \redmapper, we now go back to
the mock vs.  \redmapper\ galaxy surface density profiles shown in
Fig.~\ref{fig:dsgal}.  In the top panel, the mock galaxy density profiles of the
large and small-$\langle R^{\mathrm{mock}}_{\mathrm{mem}}\rangle$
clusters~(dotted) roughly reproduce both the shapes and amplitudes of the real
$\rzu$-split cluster profiles~(solid). If the scatter in the mock is much
larger~(smaller) than in the observations, the contrast between the galaxy
surface density profiles of the two subsamples would be much lower~(higher). The
ratio between the $\langle R^{\mathrm{mock}}_{\mathrm{mem}}\rangle$-split
profiles in the mock~($0.60$) is slightly lower than that between the
$\rzu$-split profiles in \redmapper~($0.65$) at $R/\rlam{\sim}0.1$~(dotted vs.\
solid curves in the lower panel), suggesting a similar but slightly
under-estimated scatter in the mock compared to the data.  Furthermore, the
distance at which the ratio of two $R_{\mathrm{mem}}^{\mathrm{mock}}$-split
profiles crosses unity is ${\simeq}0.3\rlam$, in excellent agreement with that
of the $\rzu$-split samples. This agreement indicates that the $0.65\rlam$
aperture we adopted for $\langle R^{\mathrm{mock}}_{\mathrm{mem}}\rangle$ is
close to the effective aperture of $\rzu$ in \redmapper. Note that since
\redmapper\ used a 2D aperture $\rlam$ instead of $r_{200m}$ to derive
$\lambda$, we boost the galaxy surface number density profiles of the mock
clusters by $\lambda/N_{\rlam}$ to match to observations in
Fig.~\ref{fig:dsgal}, where $\lambda$ and $N_{\rlam}$ are the richness
calculated from including all mock galaxies within $r_{200m}$ and the number of
mock galaxies within a cylinder of radius $\rlam$~(still within a sphere of
$r_{200m}$), respectively.

\begin{figure}
\begin{center}
    \includegraphics[width=0.5\textwidth]{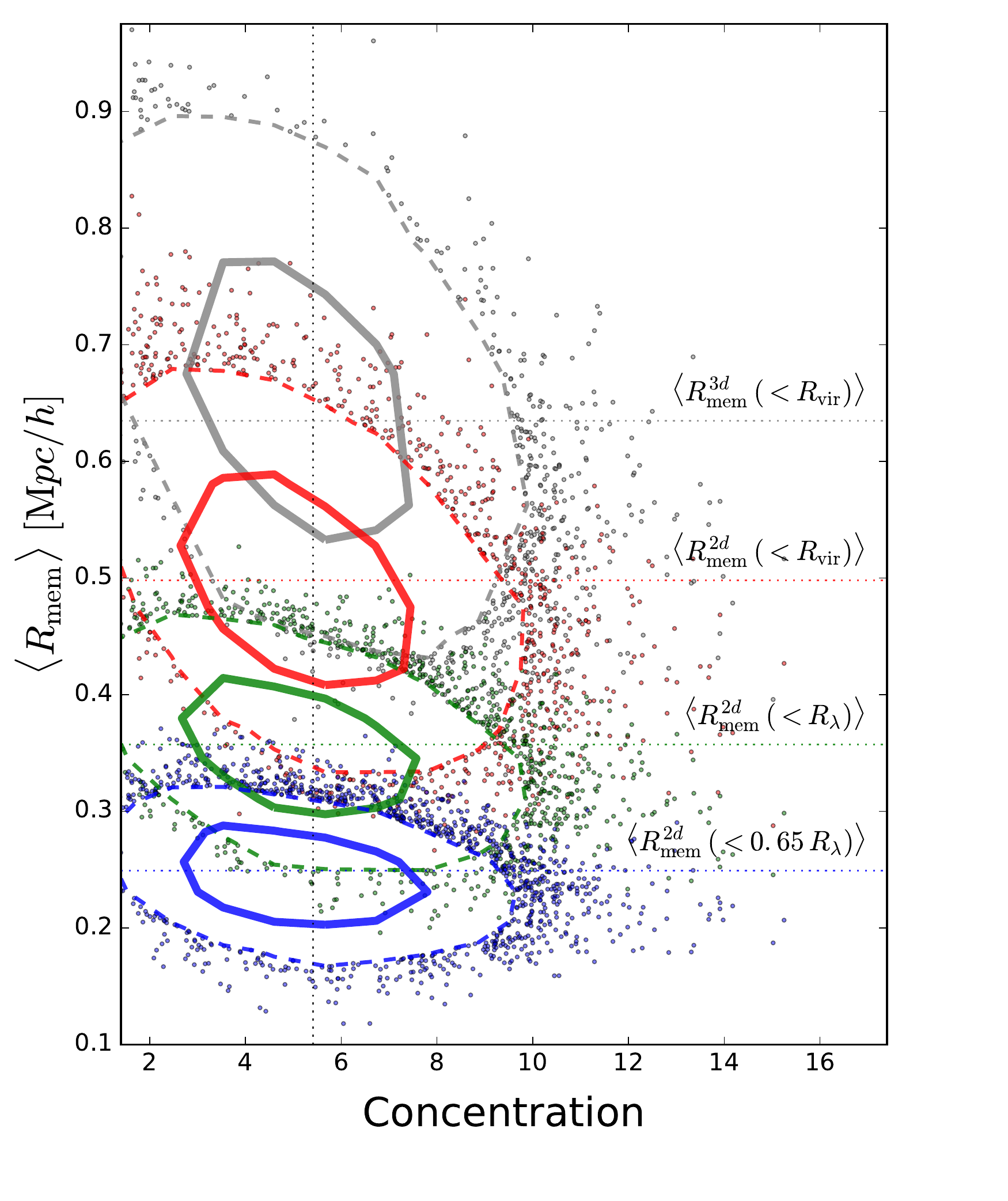}
    \caption[]{Comparison between different choices of $\rmem$ as proxies of
    halo concentration, using mock clusters with richness between $[32, 34]$ in
    the \bigmdpl\ simulation. Gray and red contours show the distributions of
    the average 3D and projected 2D membership distances, respectively,
    calculated using all galaxies within
	virial radius.  Green and blue contours are for the average 2D
	membership distance $\langle R^{2d}_{\mathrm{mem}}\rangle$ using
	galaxies projected within the \redmapper\ aperture $\rlam$ and
    $0.65 \rlam$, respectively. Horizontal dashed lines indicate the median
    values of the four $\rmem$ vs.  concentration distributions, while the
    vertical dashed line the median value of halo concentration. The correlation
    between $\rmem$ and halo concentration becomes much weaker when projected
    distances are used and fewer galaxies are included, but still persists in
    the bottom blue contour where the definition of $\rmem$ closely resembles
    the $\rzu$ measurement in \redmapper.
} \label{fig:rmemvsconc}
\end{center}
\end{figure}

With both the mock cluster and membership galaxy catalogues at hand, we first
measure several different types of $\rmem$, and compare them with the underlying
dark matter concentration in Fig.~\ref{fig:rmemvsconc}. From top to bottom, the
four contours show the distributions of four different types of $\rmem$ against
concentration $c$, including 1) the 3D average membership distance averaged over
all members within $r_{200m}$, 2) 2D average membership distance averaged over
all members within $r_{200m}$, 3) 2D average membership distance averaged over
all members within $\rlam$, and 4) 2D average membership distance averaged over
all members within $\rlam$, for mock clusters with richness between $32$ and
$34$~(the average richness of the sample is $33$). The horizontal dotted lines
indicate the median values of $\rmem$ for the four estimators, and the vertical
dotted line the median value of halo concentration. Unsurprisingly, the 3D
$\rmem$ estimator~(gray) shows the strongest correlation with $c$, with a
\texttt{Spearman}'s cross-correlation coefficient of $\rho_{cc}{=}-0.48$, where
the scatter is entirely due to stochasticity of small galaxy numbers per
cluster.  The correlation becomes slightly weaker when the 2D $\rmem$
estimator~(red; $\rho_{cc}{=}-0.45$) is used, and weakens even further when
extra aperture cuts of $\rlam$~(green; $\rho_{cc}{=}-0.42$) and
$0.65\rlam$~(blue; $\rho_{cc}{=}-0.31$) on $R$ are placed.
% -0.483881599751
% -0.449992212342
% -0.422013607874
% -0.310729696705

\begin{figure*}
\begin{center}
    \includegraphics[width=0.8\textwidth]{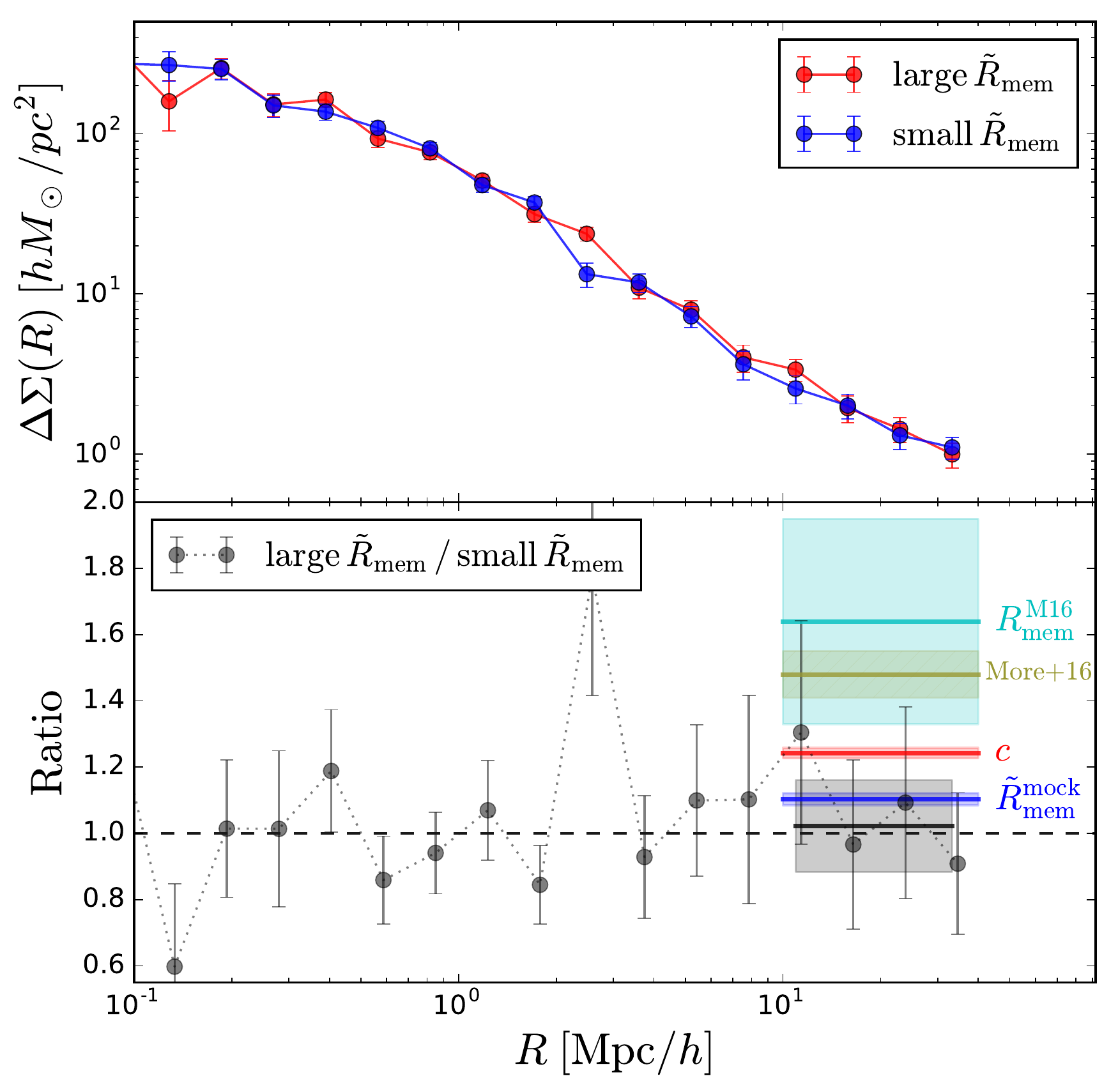}
    \caption[]{Stacked weak lensing measurements of the large- and small-$\rzu$
    cluster subsamples. The bottom panels show the ratios
	between the two profiles, and the gray band highlights
    the ratio between the two large-scale biases and its 1-$\sigma$ error,
    computed using scales above $10\,\hmpc$. Red and blue thin bands indicate
    the two types of assembly bias signals predicted by $\lcdm$, using
    subsamples split by halo concentration $c$ and
    $\tilde{R}^{\mathrm{mock}}_{\mathrm{mem}}$, respectively. Cyan and yellow
    bands indicate the bias ratios derived by M16 and \citet{more16}, which are
    $1.5\sigma$ and $3.5\sigma$ higher than the maximum assembly bias
    signal~(red band), respectively. The weak lensing measurement~(gray band) is
    consistent with the assembly bias signal predicted from using
    $\tilde{R}^{\mathrm{mock}}_{\mathrm{mem}}$, i.e., the average projected
    membership distance of member galaxies within $0.65\rlam$ in the \bigmdpl\
    mock cluster catalogue.} \label{fig:clusterwl}
\end{center}
\end{figure*}

Fortunately, there is still substantial correlation left between $\langle
R^{\mathrm{mock}}_{\mathrm{mem}}\rangle$ and $c$, and the large~(small)-$\langle
R^{\mathrm{mock}}_{\mathrm{mem}}\rangle$ subsample is still dominated by the
intrinsically low~(high)-concentration clusters. In particular, the maximum
assembly bias, as measured by the bias ratio between the low and
high-concentration subsamples in the mock catalogue~(using their 3D
auto-correlation functions above $10\hmpc$), is $1.24\pm0.02$, while
the assembly bias retained by the $\langle R^{\mathrm{mock}}_{\mathrm{mem}}\rangle$
estimator is $60\%$ weaker, i.e., $1.10\pm0.02$. However, we emphasize that this bias
ratio of $1.10\pm0.02$ derived from using $\langle
R^{\mathrm{mock}}_{\mathrm{mem}}\rangle$ should be regarded as an upper limit of
the observable signal in \redmapper\ using $\rzu$, because in the mock data the
scatter between $\rmem$ and $c$ is under-estimated.

Expecting a bias ratio at most $1.10$ based on mock data, we are now ready to go
back to the \redmapper\ clusters, and examine whether the level of halo assembly
bias revealed by $\rzu$ in the SDSS data is consistent with this $\lcdm$
expectation. Following M16, we employ the stacked weak lensing of clusters~(see
\citealt{simet16} for details) to examine the ratio between the large-scale
clustering biases of the large and small-$\rzu$ subsamples.

Fig.~\ref{fig:clusterwl} presents the key result of this paper. In the top
panel, we compare the stacked weak lensing signals between the large~(red) and
small-$\rzu$~(blue) subsamples of clusters on scales between $0.1$ and
$40\,\hmpc$. The two weak lensing profiles are consistent with each other on all
measured scales, indicating no discrepancy in either halo mass ~(${<}1\,\hmpc$)
or large-scale bias~(${>}10\,\hmpc$). The bottom panel shows the ratio between
the weak lensing signals of the large and small-$\rzu$ clusters. The black
horizontal line with gray band indicate the large-scale bias ratio and its
1-$\sigma$ uncertainty~($1.02\pm0.14$), using the weak lensing measurements
above $10\hmpc$. The observed ratio is consistent with the blue thin
band~($1.10\pm0.02$), which shows the expected levels of assembly bias derived
from the large and small-$\langle R^{\mathrm{mock}}_{\mathrm{mem}}\rangle$
clusters in the mock catalogue.  However, with large errorbars, the observed
ratio is also consistent with unity, i.e., having no assembly bias. Also shown
in the bottom panel are three bands indicating the maximum assembly bias if halo
concentration is accessible~(red), and the relatively high bias ratios measured
in M16~(cyan) and~\citet{more16}~(yellow). The strong discrepancies between
these two measurements and the maximum ratio disappear almost entirely after we
remove the projection effects from $\rmem$ by using $\rzu$.

Therefore, our re-analysis demonstrates that the level of cluster assembly bias
in the SDSS \redmapper\ catalogue is consistent with the prediction from $\lcdm$
structure formation theory. Although we do not detect halo assembly bias, we are
only limited by the statistical uncertainties in the weak lensing measurements,
rather than projection effects in the estimate of $\rmem$. This is very
encouraging because, with enough statistics, the residual assembly bias signal
can be robustly detected with little systematic uncertainties in ongoing and
future cluster surveys.

\section{Conclusion}
\label{sec:conc}

Halo assembly bias is one of the most robust features of structure formation
under $\lcdm$, but its signature in massive clusters could be affected by
systematic uncertainties due to projection effects, as well as noisy
measurements of halo concentrations $c$ using average membership distances
$\rmem$.

Using the public SDSS DR8 \redmapper\ cluster catalogue, we developed a mark
correlation statistic for diagnosing projection effects in $\rmem$,
particularly the contamination of cluster membership probabilities $p_m$ by
background galaxies. By examining the mark correlation of $\rmiya$, the $\rmem$
estimator proposed by \citet{miyatake16}, we discovered that the $\rmiya$ values
are significantly correlated between two clusters that are close to each other
on the sky but well-separated in redshift.  Therefore, the strong discrepancy
between clustering biases of two cluster subsamples split by $\rmiya$, measured
via weak lensing~\citep{miyatake16} or cross-correlation with photometric galaxy
catalogues~\citep{more16}, can be largely attributed to the background
contamination in $p_m$, rather than an anomalous signal of halo assembly bias.

To re-assess the level of cluster assembly bias in SDSS, we have developed a new
variant of the $\rmem$ estimator by excluding the low-membership probability
galaxies~($p_m{<}0.8$) that are strongly affected by the contamination in $p_m$.
Further null test indicates that the new estimator $\rzu$ is a clean indicator
of galaxy concentration free of projection effects. In the longer term, however,
it would be valuable to have a more robust method of assigning $p_m$ values that
can be used directly with minimal projection effects on both cluster richness
and $\rmem$.

In order to evaluate the impact of scatter between $\rzu$ and halo concentration
$c$ on cluster assembly bias detection, we constructed mock catalogues of
clusters and red-sequence galaxies from a large-volume $\lcdm$ simulation, and
predicted that the bias ratio between the two cluster subsamples split by $\rzu$
should be at most $1.10\pm0.02$, i.e., at least $60\%$ weaker than the signal predicted
using clusters split by halo concentration~($1.24$). This reduction is mainly
caused by the combination of Poisson fluctuations in the galaxy number profile
per cluster, and the small aperture size and projected distances used for
defining $\rzu$.

From the weak lensing measurements of \redmapper\ cluster subsamples split by
$\rzu$, we discovered that while having equal average masses, the bias ratio of
the two subsamples is $1.02\pm0.14$, consistent with the prediction from our
mock $\lcdm$ clusters as well as unity, i.e., having no assembly bias.
Therefore, although no detection can be made due to large statistical
uncertainties in the bias ratio, our result demonstrates that the level of halo
assembly bias exhibited in the SDSS \redmapper\ catalogue is consistent with
$\lcdm$.

Within the present SDSS \redmapper\ catalogue, a statistically significant
detection of cluster assembly bias requires either significant reduction in the
uncertainties of weak lensing measurements, or great improvement in the
assignment of membership probabilities. However, with an order of magnitude
increase in the observed cluster number counts, ongoing surveys like the Dark
Energy Survey~\citep{des2005} and the Hyper Suprime-Cam~\citep{miyazaki2012}
will deliver a smoking-gun detection of halo assembly bias in the very near
future.

\section*{Acknowlegements}

We thank Hironao Miyatake and Surhud More for carefully reading all previous versions of the manuscript and
for giving detailed comments and suggestions that have greatly improved the manuscript. We also thank Bhuvnesh
Jain, Neal Dalal, and Zheng Zheng for stimulating discussions at the early stage of this investigation, and
David Spergel and Masahiro Takada for helpful comments on the final version of the manuscript. YZ and RM
acknowledge the support by the U.S. Department of Energy~(DOE) Early Career Program. YZ is also supported by a
CCAPP fellowship. ER acknowledges support by the DOE Early Career Program, DOE grant DE-SC0015975, and the
Sloan Foundation, grant FG-2016-6443. ESR is supported in part by the DOE contract to SLAC no.
DE-AC02-76SF00515.

% \begin{figure}
% \begin{center}
    % \includegraphics[width=0.5\textwidth]{{figs/cluster_mgsblue}.pdf}
    % \caption[]{TBD}
% \end{center}
% \end{figure}


% \clearpage
\bibliographystyle{mnras}
% \bibliography{bibliography/biblio} %

\begin{thebibliography}{}
\makeatletter
\relax
\def\mn@urlcharsother{\let\do\@makeother \do\$\do\&\do\#\do\^\do\_\do\%\do\~}
\def\mn@doi{\begingroup\mn@urlcharsother \@ifnextchar [ {\mn@doi@}
  {\mn@doi@[]}}
\def\mn@doi@[#1]#2{\def\@tempa{#1}\ifx\@tempa\@empty \href
  {http://dx.doi.org/#2} {doi:#2}\else \href {http://dx.doi.org/#2} {#1}\fi
  \endgroup}
\def\mn@eprint#1#2{\mn@eprint@#1:#2::\@nil}
\def\mn@eprint@arXiv#1{\href {http://arxiv.org/abs/#1} {{\tt arXiv:#1}}}
\def\mn@eprint@dblp#1{\href {http://dblp.uni-trier.de/rec/bibtex/#1.xml}
  {dblp:#1}}
\def\mn@eprint@#1:#2:#3:#4\@nil{\def\@tempa {#1}\def\@tempb {#2}\def\@tempc
  {#3}\ifx \@tempc \@empty \let \@tempc \@tempb \let \@tempb \@tempa \fi \ifx
  \@tempb \@empty \def\@tempb {arXiv}\fi \@ifundefined
  {mn@eprint@\@tempb}{\@tempb:\@tempc}{\expandafter \expandafter \csname
  mn@eprint@\@tempb\endcsname \expandafter{\@tempc}}}

\bibitem[\protect\citeauthoryear{{Aihara} et~al.,}{{Aihara}
  et~al.}{2011}]{aihara2011}
{Aihara} H.,  et~al., 2011, \mn@doi [\apjs] {10.1088/0067-0049/193/2/29}, \href
  {http://adsabs.harvard.edu/abs/2011ApJS..193...29A} {193, 29}

\bibitem[\protect\citeauthoryear{{Bardeen}, {Bond}, {Kaiser}  \&
  {Szalay}}{{Bardeen} et~al.}{1986}]{bardeen1986}
{Bardeen} J.~M.,  {Bond} J.~R.,  {Kaiser} N.,   {Szalay} A.~S.,  1986, \mn@doi
  [\apj] {10.1086/164143}, \href
  {http://adsabs.harvard.edu/abs/1986ApJ...304...15B} {304, 15}

\bibitem[\protect\citeauthoryear{{Baxter}, {Rozo}, {Jain}, {Rykoff}  \&
  {Wechsler}}{{Baxter} et~al.}{2016}]{baxter16}
{Baxter} E.~J.,  {Rozo} E.,  {Jain} B.,  {Rykoff} E.,   {Wechsler} R.~H.,
  2016, preprint, \href {http://adsabs.harvard.edu/abs/2016arXiv160400048B} {}
  (\mn@eprint {arXiv} {1604.00048})

\bibitem[\protect\citeauthoryear{{Behroozi}, {Wechsler}  \& {Wu}}{{Behroozi}
  et~al.}{2013}]{behroozi13}
{Behroozi} P.~S.,  {Wechsler} R.~H.,   {Wu} H.-Y.,  2013, \mn@doi [\apj]
  {10.1088/0004-637X/762/2/109}, \href
  {http://adsabs.harvard.edu/abs/2013ApJ...762..109B} {762, 109}

\bibitem[\protect\citeauthoryear{{Beisbart} \& {Kerscher}}{{Beisbart} \&
  {Kerscher}}{2000}]{beisbart2000}
{Beisbart} C.,  {Kerscher} M.,  2000, \mn@doi [\apj] {10.1086/317788}, \href
  {http://adsabs.harvard.edu/abs/2000ApJ...545....6B} {545, 6}

% \bibitem[\protect\citeauthoryear{{Berlind}, {Kazin}, {Blanton}, {Pueblas},
  % {Scoccimarro}  \& {Hogg}}{{Berlind} et~al.}{2006}]{berlind2006}
% {Berlind} A.~A.,  {Kazin} E.,  {Blanton} M.~R.,  {Pueblas} S.,  {Scoccimarro}
  % R.,   {Hogg} D.~W.,  2006, ArXiv Astrophysics e-prints, \href
  % {http://adsabs.harvard.edu/abs/2006astro.ph.10524B} {}

\bibitem[Berlind et al.(2006)]{berlind2006} Berlind, A.~A., Kazin, E., Blanton, M.~R., et al.\ 2006, arXiv:astro-ph/0610524



\bibitem[\protect\citeauthoryear{{Blanton} \& {Berlind}}{{Blanton} \&
  {Berlind}}{2007}]{blanton2007}
{Blanton} M.~R.,  {Berlind} A.~A.,  2007, \mn@doi [\apj] {10.1086/512478},
  \href {http://adsabs.harvard.edu/abs/2007ApJ...664..791B} {664, 791}

\bibitem[\protect\citeauthoryear{{Budzynski}, {Koposov}, {McCarthy}, {McGee}
  \& {Belokurov}}{{Budzynski} et~al.}{2012}]{budzynski2012}
{Budzynski} J.~M.,  {Koposov} S.~E.,  {McCarthy} I.~G.,  {McGee} S.~L.,
  {Belokurov} V.,  2012, \mn@doi [\mnras] {10.1111/j.1365-2966.2012.20663.x},
  \href {http://adsabs.harvard.edu/abs/2012MNRAS.423..104B} {423, 104}

\bibitem[\protect\citeauthoryear{{Campbell}, {van den Bosch}, {Hearin},
  {Padmanabhan}, {Berlind}, {Mo}, {Tinker}  \& {Yang}}{{Campbell}
  et~al.}{2015}]{campbell2015}
{Campbell} D.,  {van den Bosch} F.~C.,  {Hearin} A.,  {Padmanabhan} N.,
  {Berlind} A.,  {Mo} H.~J.,  {Tinker} J.,   {Yang} X.,  2015, \mn@doi [\mnras]
  {10.1093/mnras/stv1091}, \href
  {http://adsabs.harvard.edu/abs/2015MNRAS.452..444C} {452, 444}

\bibitem[\protect\citeauthoryear{{Castignani} \& {Benoist}}{{Castignani} \&
  {Benoist}}{2016}]{castignani2016}
{Castignani} G.,  {Benoist} C.,  2016, preprint, \href
  {http://adsabs.harvard.edu/abs/2016arXiv160608744C} {} (\mn@eprint {arXiv}
  {1606.08744})

\bibitem[\protect\citeauthoryear{{Chen}, {Kravtsov}, {Prada}, {Sheldon},
  {Klypin}, {Blanton}, {Brinkmann}  \& {Thakar}}{{Chen}
  et~al.}{2006}]{chen2006}
{Chen} J.,  {Kravtsov} A.~V.,  {Prada} F.,  {Sheldon} E.~S.,  {Klypin} A.~A.,
  {Blanton} M.~R.,  {Brinkmann} J.,   {Thakar} A.~R.,  2006, \mn@doi [\apj]
  {10.1086/504462}, \href {http://adsabs.harvard.edu/abs/2006ApJ...647...86C}
  {647, 86}

\bibitem[\protect\citeauthoryear{{Croton}, {Gao}  \& {White}}{{Croton}
  et~al.}{2007}]{croton2007}
{Croton} D.~J.,  {Gao} L.,   {White} S.~D.~M.,  2007, \mn@doi [\mnras]
  {10.1111/j.1365-2966.2006.11230.x}, \href
  {http://adsabs.harvard.edu/abs/2007MNRAS.374.1303C} {374, 1303}

\bibitem[\protect\citeauthoryear{{Dalal}}{{Dalal}}{2016}]{dalal2016}
{Dalal} N.,  2016, \mn@doi [Physics Online Journal] {10.1103/Physics.9.9},
  \href {http://adsabs.harvard.edu/abs/2016PhyOJ...9....9D} {9, 9}

\bibitem[\protect\citeauthoryear{{Dalal}, {White}, {Bond}  \&
  {Shirokov}}{{Dalal} et~al.}{2008}]{dalal08}
{Dalal} N.,  {White} M.,  {Bond} J.~R.,   {Shirokov} A.,  2008, \mn@doi [\apj]
  {10.1086/591512}, \href {http://adsabs.harvard.edu/abs/2008ApJ...687...12D}
  {687, 12}

\bibitem[\protect\citeauthoryear{{Deason}, {Conroy}, {Wetzel}  \&
  {Tinker}}{{Deason} et~al.}{2013}]{deason2013}
{Deason} A.~J.,  {Conroy} C.,  {Wetzel} A.~R.,   {Tinker} J.~L.,  2013, \mn@doi
  [\apj] {10.1088/0004-637X/777/2/154}, \href
  {http://adsabs.harvard.edu/abs/2013ApJ...777..154D} {777, 154}

\bibitem[\protect\citeauthoryear{{Diemand}, {Moore}  \& {Stadel}}{{Diemand}
  et~al.}{2004}]{diemand2004}
{Diemand} J.,  {Moore} B.,   {Stadel} J.,  2004, \mn@doi [\mnras]
  {10.1111/j.1365-2966.2004.07940.x}, \href
  {http://adsabs.harvard.edu/abs/2004MNRAS.352..535D} {352, 535}

\bibitem[\protect\citeauthoryear{{Diemand}, {Kuhlen}  \& {Madau}}{{Diemand}
  et~al.}{2007}]{diemand2007}
{Diemand} J.,  {Kuhlen} M.,   {Madau} P.,  2007, \mn@doi [\apj]
  {10.1086/520573}, \href {http://adsabs.harvard.edu/abs/2007ApJ...667..859D}
  {667, 859}

\bibitem[\protect\citeauthoryear{{Erickson}, {Cunha}  \& {Evrard}}{{Erickson}
  et~al.}{2011}]{erickson2011}
{Erickson} B.~M.~S.,  {Cunha} C.~E.,   {Evrard} A.~E.,  2011, \mn@doi [\prd]
  {10.1103/PhysRevD.84.103506}, \href
  {http://adsabs.harvard.edu/abs/2011PhRvD..84j3506E} {84, 103506}

\bibitem[\protect\citeauthoryear{{Gao} \& {White}}{{Gao} \&
  {White}}{2007}]{gao2007}
{Gao} L.,  {White} S.~D.~M.,  2007, \mn@doi [\mnras]
  {10.1111/j.1745-3933.2007.00292.x}, \href
  {http://adsabs.harvard.edu/abs/2007MNRAS.377L...5G} {377, L5}

\bibitem[\protect\citeauthoryear{{Gao}, {White}, {Jenkins}, {Stoehr}  \&
  {Springel}}{{Gao} et~al.}{2004}]{gao2004}
{Gao} L.,  {White} S.~D.~M.,  {Jenkins} A.,  {Stoehr} F.,   {Springel} V.,
  2004, \mn@doi [\mnras] {10.1111/j.1365-2966.2004.08360.x}, \href
  {http://adsabs.harvard.edu/abs/2004MNRAS.355..819G} {355, 819}

\bibitem[\protect\citeauthoryear{{Gao}, {Springel}  \& {White}}{{Gao}
  et~al.}{2005}]{gao2005}
{Gao} L.,  {Springel} V.,   {White} S.~D.~M.,  2005, \mn@doi [\mnras]
  {10.1111/j.1745-3933.2005.00084.x}, \href
  {http://adsabs.harvard.edu/abs/2005MNRAS.363L..66G} {363, L66}

\bibitem[\protect\citeauthoryear{{Hahn}, {Porciani}, {Dekel}  \&
  {Carollo}}{{Hahn} et~al.}{2009}]{hahn2009}
{Hahn} O.,  {Porciani} C.,  {Dekel} A.,   {Carollo} C.~M.,  2009, \mn@doi
  [\mnras] {10.1111/j.1365-2966.2009.15271.x}, \href
  {http://adsabs.harvard.edu/abs/2009MNRAS.398.1742H} {398, 1742}

\bibitem[\protect\citeauthoryear{{Han}, {Cole}, {Frenk}  \& {Jing}}{{Han}
  et~al.}{2016}]{han2016}
{Han} J.,  {Cole} S.,  {Frenk} C.~S.,   {Jing} Y.,  2016, \mn@doi [\mnras]
  {10.1093/mnras/stv2900}, \href
  {http://adsabs.harvard.edu/abs/2016MNRAS.457.1208H} {457, 1208}

\bibitem[\protect\citeauthoryear{{Harker}, {Cole}, {Helly}, {Frenk}  \&
  {Jenkins}}{{Harker} et~al.}{2006}]{harker2006}
{Harker} G.,  {Cole} S.,  {Helly} J.,  {Frenk} C.,   {Jenkins} A.,  2006,
  \mn@doi [\mnras] {10.1111/j.1365-2966.2006.10022.x}, \href
  {http://adsabs.harvard.edu/abs/2006MNRAS.367.1039H} {367, 1039}

\bibitem[\protect\citeauthoryear{{Jing}, {Suto}  \& {Mo}}{{Jing}
  et~al.}{2007}]{jing2007}
{Jing} Y.~P.,  {Suto} Y.,   {Mo} H.~J.,  2007, \mn@doi [\apj] {10.1086/511130},
  \href {http://adsabs.harvard.edu/abs/2007ApJ...657..664J} {657, 664}

\bibitem[\protect\citeauthoryear{{Kauffmann}, {Li}, {Zhang}  \&
  {Weinmann}}{{Kauffmann} et~al.}{2013}]{kauffmann2013}
{Kauffmann} G.,  {Li} C.,  {Zhang} W.,   {Weinmann} S.,  2013, \mn@doi [\mnras]
  {10.1093/mnras/stt007}, \href
  {http://adsabs.harvard.edu/abs/2013MNRAS.430.1447K} {430, 1447}

\bibitem[\protect\citeauthoryear{{Klypin}, {Yepes}, {Gottl{\"o}ber}, {Prada}
  \& {He{\ss}}}{{Klypin} et~al.}{2016}]{klypin16}
{Klypin} A.,  {Yepes} G.,  {Gottl{\"o}ber} S.,  {Prada} F.,   {He{\ss}} S.,
  2016, \mn@doi [\mnras] {10.1093/mnras/stw248}, \href
  {http://adsabs.harvard.edu/abs/2016MNRAS.457.4340K} {457, 4340}

\bibitem[\protect\citeauthoryear{{Lacerna}, {Padilla}  \&
  {Stasyszyn}}{{Lacerna} et~al.}{2014}]{lacerna2014}
{Lacerna} I.,  {Padilla} N.,   {Stasyszyn} F.,  2014, \mn@doi [\mnras]
  {10.1093/mnras/stu1318}, \href
  {http://adsabs.harvard.edu/abs/2014MNRAS.443.3107L} {443, 3107}

\bibitem[\protect\citeauthoryear{{Lehmann}, {Mao}, {Becker}, {Skillman}  \&
  {Wechsler}}{{Lehmann} et~al.}{2015}]{lehmann2015}
{Lehmann} B.~V.,  {Mao} Y.-Y.,  {Becker} M.~R.,  {Skillman} S.~W.,   {Wechsler}
  R.~H.,  2015, preprint, \href
  {http://adsabs.harvard.edu/abs/2015arXiv151005651L} {} (\mn@eprint {arXiv}
  {1510.05651})

\bibitem[\protect\citeauthoryear{{Li}, {Mo}  \& {Gao}}{{Li}
  et~al.}{2008}]{li2008}
{Li} Y.,  {Mo} H.~J.,   {Gao} L.,  2008, \mn@doi [\mnras]
  {10.1111/j.1365-2966.2008.13667.x}, \href
  {http://adsabs.harvard.edu/abs/2008MNRAS.389.1419L} {389, 1419}

\bibitem[\protect\citeauthoryear{{Lin}, {Mandelbaum}, {Huang}, {Huang},
  {Dalal}, {Diemer}, {Jian}  \& {Kravtsov}}{{Lin} et~al.}{2016}]{lin16}
{Lin} Y.-T.,  {Mandelbaum} R.,  {Huang} Y.-H.,  {Huang} H.-J.,  {Dalal} N.,
  {Diemer} B.,  {Jian} H.-Y.,   {Kravtsov} A.,  2016, \mn@doi [\apj]
  {10.3847/0004-637X/819/2/119}, \href
  {http://adsabs.harvard.edu/abs/2016ApJ...819..119L} {819, 119}

\bibitem[\protect\citeauthoryear{{Ludlow}, {Navarro}, {Springel}, {Jenkins},
  {Frenk}  \& {Helmi}}{{Ludlow} et~al.}{2009}]{ludlow2009}
{Ludlow} A.~D.,  {Navarro} J.~F.,  {Springel} V.,  {Jenkins} A.,  {Frenk}
  C.~S.,   {Helmi} A.,  2009, \mn@doi [\apj] {10.1088/0004-637X/692/1/931},
  \href {http://adsabs.harvard.edu/abs/2009ApJ...692..931L} {692, 931}

\bibitem[\protect\citeauthoryear{{McEwen} \& {Weinberg}}{{McEwen} \&
  {Weinberg}}{2016}]{mcewen2016}
{McEwen} J.~E.,  {Weinberg} D.~H.,  2016, preprint, \href
  {http://adsabs.harvard.edu/abs/2016arXiv160102693M} {} (\mn@eprint {arXiv}
  {1601.02693})

\bibitem[\protect\citeauthoryear{{Medezinski}, {Battaglia}, {Coupon}, {Cen},
  {Gaspari}, {Strauss}  \& {Spergel}}{{Medezinski}
  et~al.}{2016}]{medezinski2016}
{Medezinski} E.,  {Battaglia} N.,  {Coupon} J.,  {Cen} R.,  {Gaspari} M.,
  {Strauss} M.~A.,   {Spergel} D.~N.,  2016, preprint, \href
  {http://adsabs.harvard.edu/abs/2016arXiv161001624M} {} (\mn@eprint {arXiv}
  {1610.01624})

\bibitem[\protect\citeauthoryear{{Miyatake}, {More}, {Takada}, {Spergel},
  {Mandelbaum}, {Rykoff}  \& {Rozo}}{{Miyatake} et~al.}{2016}]{miyatake16}
{Miyatake} H.,  {More} S.,  {Takada} M.,  {Spergel} D.~N.,  {Mandelbaum} R.,
  {Rykoff} E.~S.,   {Rozo} E.,  2016, \mn@doi [Physical Review Letters]
  {10.1103/PhysRevLett.116.041301}, \href
  {http://adsabs.harvard.edu/abs/2016PhRvL.116d1301M} {116, 041301}

\bibitem[\protect\citeauthoryear{{Miyazaki} et~al.,}{{Miyazaki}
  et~al.}{2012}]{miyazaki2012}
{Miyazaki} S.,  et~al., 2012, in Ground-based and Airborne Instrumentation for
  Astronomy IV. p. 84460Z, \mn@doi{10.1117/12.926844}

\bibitem[\protect\citeauthoryear{{More} et~al.,}{{More} et~al.}{2016}]{more16}
{More} S.,  et~al., 2016, \mn@doi [\apj] {10.3847/0004-637X/825/1/39}, \href
  {http://adsabs.harvard.edu/abs/2016ApJ...825...39M} {825, 39}

\bibitem[\protect\citeauthoryear{{Nagai} \& {Kravtsov}}{{Nagai} \&
  {Kravtsov}}{2005}]{nagai2005}
{Nagai} D.,  {Kravtsov} A.~V.,  2005, \mn@doi [\apj] {10.1086/426016}, \href
  {http://adsabs.harvard.edu/abs/2005ApJ...618..557N} {618, 557}

\bibitem[\protect\citeauthoryear{{Noh} \& {Cohn}}{{Noh} \&
  {Cohn}}{2012}]{noh2012}
{Noh} Y.,  {Cohn} J.~D.,  2012, \mn@doi [\mnras]
  {10.1111/j.1365-2966.2012.21810.x}, \href
  {http://adsabs.harvard.edu/abs/2012MNRAS.426.1829N} {426, 1829}

\bibitem[\protect\citeauthoryear{{Paranjape}, {Kova{\v c}}, {Hartley}  \&
  {Pahwa}}{{Paranjape} et~al.}{2015}]{paranjape2015}
{Paranjape} A.,  {Kova{\v c}} K.,  {Hartley} W.~G.,   {Pahwa} I.,  2015,
  \mn@doi [\mnras] {10.1093/mnras/stv2137}, \href
  {http://adsabs.harvard.edu/abs/2015MNRAS.454.3030P} {454, 3030}

\bibitem[\protect\citeauthoryear{{Planck Collaboration} et~al.,}{{Planck
  Collaboration} et~al.}{2015}]{planck15}
{Planck Collaboration} et~al., 2015, preprint, \href
  {http://adsabs.harvard.edu/abs/2015arXiv150201589P} {} (\mn@eprint {arXiv}
  {1502.01589})

\bibitem[\protect\citeauthoryear{{Rykoff} et~al.,}{{Rykoff}
  et~al.}{2014}]{rykoff14}
{Rykoff} E.~S.,  et~al., 2014, \mn@doi [\apj] {10.1088/0004-637X/785/2/104},
  \href {http://adsabs.harvard.edu/abs/2014ApJ...785..104R} {785, 104}

\bibitem[\protect\citeauthoryear{{Sheth} \& {Tormen}}{{Sheth} \&
  {Tormen}}{1999}]{sheth1999}
{Sheth} R.~K.,  {Tormen} G.,  1999, \mn@doi [\mnras]
  {10.1046/j.1365-8711.1999.02692.x}, \href
  {http://adsabs.harvard.edu/abs/1999MNRAS.308..119S} {308, 119}

\bibitem[\protect\citeauthoryear{{Sheth} \& {Tormen}}{{Sheth} \&
  {Tormen}}{2004}]{sheth2004}
{Sheth} R.~K.,  {Tormen} G.,  2004, \mn@doi [\mnras]
  {10.1111/j.1365-2966.2004.07733.x}, \href
  {http://adsabs.harvard.edu/abs/2004MNRAS.350.1385S} {350, 1385}

% \bibitem[\protect\citeauthoryear{{Sheth}, {Connolly}  \& {Skibba}}{{Sheth}
  % et~al.}{2005}]{sheth2005}
% {Sheth} R.~K.,  {Connolly} A.~J.,   {Skibba} R.,  2005, ArXiv Astrophysics
  % e-prints, \href {http://adsabs.harvard.edu/abs/2005astro.ph.11773S} {}

\bibitem[Sheth et al.(2005)]{sheth2005} Sheth, R.~K., Connolly, A.~J., \& Skibba, R.\ 2005, arXiv:astro-ph/0511773



\bibitem[\protect\citeauthoryear{{Simet}, {McClintock}, {Mandelbaum}, {Rozo},
  {Rykoff}, {Sheldon}  \& {Wechsler}}{{Simet} et~al.}{2016}]{simet16}
{Simet} M.,  {McClintock} T.,  {Mandelbaum} R.,  {Rozo} E.,  {Rykoff} E.,
  {Sheldon} E.,   {Wechsler} R.~H.,  2016, preprint, \href
  {http://adsabs.harvard.edu/abs/2016arXiv160306953S} {} (\mn@eprint {arXiv}
  {1603.06953})

\bibitem[\protect\citeauthoryear{{Skibba}, {Sheth}, {Connolly}  \&
  {Scranton}}{{Skibba} et~al.}{2006}]{skibba06}
{Skibba} R.,  {Sheth} R.~K.,  {Connolly} A.~J.,   {Scranton} R.,  2006, \mn@doi
  [\mnras] {10.1111/j.1365-2966.2006.10196.x}, \href
  {http://adsabs.harvard.edu/abs/2006MNRAS.369...68S} {369, 68}

\bibitem[\protect\citeauthoryear{Stoyan \& Stoyan}{Stoyan \&
  Stoyan}{1994}]{stoyan94}
Stoyan D.,  Stoyan H.,  1994, Fractals, Random Shapes, and Point Fields:
  Methods of Geometrical Statistics.
John Wiley \& Sons

\bibitem[\protect\citeauthoryear{{Sunayama}, {Hearin}, {Padmanabhan}  \&
  {Leauthaud}}{{Sunayama} et~al.}{2016}]{sunayama2016}
{Sunayama} T.,  {Hearin} A.~P.,  {Padmanabhan} N.,   {Leauthaud} A.,  2016,
  \mn@doi [\mnras] {10.1093/mnras/stw332}, \href
  {http://adsabs.harvard.edu/abs/2016MNRAS.458.1510S} {458, 1510}

% \bibitem[\protect\citeauthoryear{{The Dark Energy Survey Collaboration}}{{The
  % Dark Energy Survey Collaboration}}{2005}]{des2005}
% {The Dark Energy Survey Collaboration} 2005, ArXiv Astrophysics e-prints, \href
  % {http://adsabs.harvard.edu/abs/2005astro.ph.10346T} {}

  \bibitem[The Dark Energy Survey Collaboration(2005)]{des2005} The Dark Energy Survey Collaboration 2005, arXiv:astro-ph/0510346

\bibitem[\protect\citeauthoryear{{Wang}, {Mo}  \& {Jing}}{{Wang}
  et~al.}{2009}]{wang2009}
{Wang} H.,  {Mo} H.~J.,   {Jing} Y.~P.,  2009, \mn@doi [\mnras]
  {10.1111/j.1365-2966.2009.14884.x}, \href
  {http://adsabs.harvard.edu/abs/2009MNRAS.396.2249W} {396, 2249}

\bibitem[\protect\citeauthoryear{{Wang}, {Weinmann}, {De Lucia}  \&
  {Yang}}{{Wang} et~al.}{2013}]{wang2013}
{Wang} L.,  {Weinmann} S.~M.,  {De Lucia} G.,   {Yang} X.,  2013, \mn@doi
  [\mnras] {10.1093/mnras/stt743}, \href
  {http://adsabs.harvard.edu/abs/2013MNRAS.433..515W} {433, 515}

\bibitem[\protect\citeauthoryear{{Wang}, {Sales}, {Henriques}  \&
  {White}}{{Wang} et~al.}{2014}]{wang2014}
{Wang} W.,  {Sales} L.~V.,  {Henriques} B.~M.~B.,   {White} S.~D.~M.,  2014,
  \mn@doi [\mnras] {10.1093/mnras/stu988}, \href
  {http://adsabs.harvard.edu/abs/2014MNRAS.442.1363W} {442, 1363}

\bibitem[\protect\citeauthoryear{{Wechsler}, {Zentner}, {Bullock}, {Kravtsov}
  \& {Allgood}}{{Wechsler} et~al.}{2006}]{wechsler06}
{Wechsler} R.~H.,  {Zentner} A.~R.,  {Bullock} J.~S.,  {Kravtsov} A.~V.,
  {Allgood} B.,  2006, \mn@doi [\apj] {10.1086/507120}, \href
  {http://adsabs.harvard.edu/abs/2006ApJ...652...71W} {652, 71}

\bibitem[\protect\citeauthoryear{{Weinmann}, {van den Bosch}, {Yang}  \&
  {Mo}}{{Weinmann} et~al.}{2006}]{weinmann2006}
{Weinmann} S.~M.,  {van den Bosch} F.~C.,  {Yang} X.,   {Mo} H.~J.,  2006,
  \mn@doi [\mnras] {10.1111/j.1365-2966.2005.09865.x}, \href
  {http://adsabs.harvard.edu/abs/2006MNRAS.366....2W} {366, 2}

\bibitem[\protect\citeauthoryear{{Wu}, {Rozo}  \& {Wechsler}}{{Wu}
  et~al.}{2008}]{wu2008}
{Wu} H.-Y.,  {Rozo} E.,   {Wechsler} R.~H.,  2008, \mn@doi [\apj]
  {10.1086/591929}, \href {http://adsabs.harvard.edu/abs/2008ApJ...688..729W}
  {688, 729}

\bibitem[\protect\citeauthoryear{{Yang}, {Mo}, {van den Bosch}, {Weinmann},
  {Li}  \& {Jing}}{{Yang} et~al.}{2005}]{yang2005}
{Yang} X.,  {Mo} H.~J.,  {van den Bosch} F.~C.,  {Weinmann} S.~M.,  {Li} C.,
  {Jing} Y.~P.,  2005, \mn@doi [\mnras] {10.1111/j.1365-2966.2005.09351.x},
  \href {http://adsabs.harvard.edu/abs/2005MNRAS.362..711Y} {362, 711}

\bibitem[\protect\citeauthoryear{{Yang}, {Mo}  \& {van den Bosch}}{{Yang}
  et~al.}{2006}]{yang2006}
{Yang} X.,  {Mo} H.~J.,   {van den Bosch} F.~C.,  2006, \mn@doi [\apjl]
  {10.1086/501069}, \href {http://adsabs.harvard.edu/abs/2006ApJ...638L..55Y}
  {638, L55}

\bibitem[\protect\citeauthoryear{{Yang}, {Mo}, {van den Bosch}, {Pasquali},
  {Li}  \& {Barden}}{{Yang} et~al.}{2007}]{yang2007}
{Yang} X.,  {Mo} H.~J.,  {van den Bosch} F.~C.,  {Pasquali} A.,  {Li} C.,
  {Barden} M.,  2007, \mn@doi [\apj] {10.1086/522027}, \href
  {http://adsabs.harvard.edu/abs/2007ApJ...671..153Y} {671, 153}

\bibitem[\protect\citeauthoryear{{York} et~al.,}{{York}
  et~al.}{2000}]{york2000}
{York} D.~G.,  et~al., 2000, \mn@doi [\aj] {10.1086/301513}, \href
  {http://adsabs.harvard.edu/abs/2000AJ....120.1579Y} {120, 1579}

\bibitem[\protect\citeauthoryear{{Zenteno} et~al.,}{{Zenteno}
  et~al.}{2016}]{zenteno2016}
{Zenteno} A.,  et~al., 2016, \mn@doi [\mnras] {10.1093/mnras/stw1649}, \href
  {http://adsabs.harvard.edu/abs/2016MNRAS.462..830Z} {462, 830}

\bibitem[\protect\citeauthoryear{{Zentner}, {Hearin}  \& {van den
  Bosch}}{{Zentner} et~al.}{2014}]{zentner2014}
{Zentner} A.~R.,  {Hearin} A.~P.,   {van den Bosch} F.~C.,  2014, \mn@doi
  [\mnras] {10.1093/mnras/stu1383}, \href
  {http://adsabs.harvard.edu/abs/2014MNRAS.443.3044Z} {443, 3044}

\bibitem[\protect\citeauthoryear{{Zentner}, {Hearin}, {van den Bosch}, {Lange}
  \& {Villarreal}}{{Zentner} et~al.}{2016}]{zentner2016}
{Zentner} A.~R.,  {Hearin} A.,  {van den Bosch} F.~C.,  {Lange} J.~U.,
  {Villarreal} A.,  2016, preprint, \href
  {http://adsabs.harvard.edu/abs/2016arXiv160607817Z} {} (\mn@eprint {arXiv}
  {1606.07817})

\bibitem[\protect\citeauthoryear{{Zhu}, {Zheng}, {Lin}, {Jing}, {Kang}  \&
  {Gao}}{{Zhu} et~al.}{2006}]{zhu2006}
{Zhu} G.,  {Zheng} Z.,  {Lin} W.~P.,  {Jing} Y.~P.,  {Kang} X.,   {Gao} L.,
  2006, \mn@doi [\apjl] {10.1086/501501}, \href
  {http://adsabs.harvard.edu/abs/2006ApJ...639L...5Z} {639, L5}

\bibitem[\protect\citeauthoryear{{Zu} \& {Mandelbaum}}{{Zu} \&
  {Mandelbaum}}{2015}]{zu15}
{Zu} Y.,  {Mandelbaum} R.,  2015, \mn@doi [\mnras] {10.1093/mnras/stv2062},
  \href {http://adsabs.harvard.edu/abs/2015MNRAS.454.1161Z} {454, 1161}

\bibitem[\protect\citeauthoryear{{Zu} \& {Mandelbaum}}{{Zu} \&
  {Mandelbaum}}{2016}]{zu16}
{Zu} Y.,  {Mandelbaum} R.,  2016, \mn@doi [\mnras] {10.1093/mnras/stw221},
  \href {http://adsabs.harvard.edu/abs/2016MNRAS.457.4360Z} {457, 4360}

\bibitem[\protect\citeauthoryear{{Zu} \& {Mandelbaum}}{{Zu} \&
  {Mandelbaum}}{2017}]{zu17} {Zu} Y.,  {Mandelbaum} R.,  2017, arXiv:1703.09219

\bibitem[\protect\citeauthoryear{{Zu}, {Zheng}, {Zhu}  \& {Jing}}{{Zu}
  et~al.}{2008}]{zu2008}
{Zu} Y.,  {Zheng} Z.,  {Zhu} G.,   {Jing} Y.~P.,  2008, \mn@doi [\apj]
  {10.1086/591071}, \href {http://adsabs.harvard.edu/abs/2008ApJ...686...41Z}
  {686, 41}



\bibitem[\protect\citeauthoryear{{van den Bosch}, {Jiang}, {Campbell}  \&
  {Behroozi}}{{van den Bosch} et~al.}{2016}]{vdbosch2016}
{van den Bosch} F.~C.,  {Jiang} F.,  {Campbell} D.,   {Behroozi} P.,  2016,
  \mn@doi [\mnras] {10.1093/mnras/stv2338}, \href
  {http://adsabs.harvard.edu/abs/2016MNRAS.455..158V} {455, 158}

\makeatother
\end{thebibliography}

% \begin{thebibliography}{99}
% \bibitem[\protect\citeauthoryear{Author}{2012}]{Author2012}
% Author A.~N., 2013, Journal of Improbable Astronomy, 1, 1
% \bibitem[\protect\citeauthoryear{Others}{2013}]{Others2013}
% Others S., 2012, Journal of Interesting Stuff, 17, 198
% \end{thebibliography}

%%%%%%%%%%%%%%%%%%%%%%%%%%%%%%%%%%%%%%%%%%%%%%%%%%

% Don't change these lines
\bsp	% typesetting comment
\label{lastpage}

\end{document}